\documentclass[12pt]{article}
\usepackage[left=3.5cm, right=2.5cm]{geometry}
\usepackage[english]{babel}
\usepackage[utf8]{inputenc}
\UseRawInputEncoding
\usepackage{adjustbox}
\usepackage{textcomp}
\usepackage{graphicx}
\usepackage{amsmath, amssymb, amsfonts, amsthm}
\usepackage{caption}
\usepackage{subcaption}
\usepackage{bbm}
\usepackage{enumerate}
\usepackage{array}
\usepackage{calc}
\usepackage{float}
\usepackage{setspace}
\usepackage{enumitem}
\usepackage{mathpazo}
\usepackage{booktabs}
\usepackage{colortbl}
\usepackage{multirow}
\usepackage{tabularx}
\usepackage{marginfix}
\usepackage[hyperfootnotes=false]{hyperref}
\hypersetup{
    colorlinks,
    citecolor=blue,
    filecolor=blue,
    linkcolor=blue,
    urlcolor=blue
}
\usepackage{cleveref}
\usepackage{marginnote}
\usepackage{caption}
 \usepackage{natbib}
\usepackage{newfloat}
\usepackage{url}
\usepackage{tikz,pgf}
\usetikzlibrary{arrows, automata}
\usepackage{attrib}
\usepackage{array}
\usepackage{scrextend}
\usepackage{authblk}
\usepackage{csquotes}
\usepackage{lscape}
\usepackage[toc,page]{appendix}
\usepackage{chngcntr}
\usepackage{longtable}
\usepackage{ragged2e}
\usepackage{rotating}
\usepackage{soul}

\title{Do News and Social Media Tell the Same Story? Constructing and Comparing Sentiment Spillover Networks.}

\author[1]{Fan Wu \thanks{Corresponding Author: WuF16@cardiff.ac.uk}} 
\author[1]{Anqi Liu}
\author[1]{Jing Chen}
\author[2]{Yuhua Li}
\affil[1]{Cardiff University, School of Mathematics, CF24 4AG, Cardiff, UK}
\affil[2]{Cardiff University, School of Computer Science and Informatics, CF24 4AG,
Cardiff, UK}
\date{January 2026}
\begin{document}
\maketitle
\begin{abstract}
Investor sentiment reflects the collective attitude of investors towards the asset, whether positive, negative or neutral. Market information, such as news and relevant social media posts, plays a significant role in shaping investor sentiment, which influences investment decisions accordingly. The sentiment for one single company may spill over to other relevant companies which are in the same industry. The information spillover network pattern between news and social media may also differ, as they are two different media sources. 
In this study, we introduce a network-based transfer entropy method to measure and compare the information transmission of news and social media sentiment across the technology companies. We examine whether and to what extent sentiment information from one company can transfer to other companies, and how different the spillover effect is for news and social media.  The result signifies a stronger intensity of news information flow among the tech companies after COVID-19. We also highlight the companies which act as information hubs in the sentiment network. Furthermore, we identify the companies which lead the strongest information flow chain. Overall, this study provides a novel perspective in modelling sentiment spillover under two different media sources, and we find that news and social media show a different information transmission pattern during the studied period.
\\

\textbf{Keywords:} Sentiment Spillover, Transfer Entropy, Network Modelling, Technology Industry, News and Social Media
\end{abstract}
\section{Introduction}
Asset prices are not only driven by their fundamentals, but also by investors' responses to the available market information, according to substantial evidence in the behavioural finance literature. 
Investor sentiment, or market sentiment, quantifies the attitude of investors towards the financial asset \citep{gan_sensitivity_2020}, which can be measured from positive to negative. Tracing back to \cite{keynes_general_1936}, who highlighted the impact of investor sentiment in decision making, particularly during the great depression, it was argued that emotions like optimism and pessimism could perhaps lead to speculative bubbles. The most recent example of SVB collapse was documented as the first ``Twitter bank run''\citep{BloombergLaw_SocialMediaBankRuns_2023}. The negative posts on Twitter contributed to the depositors withdrawing around $20\%$ of its total deposits (roughly 32 billion dollars) in a few hours \citep{cookson_social_2026}.  A number of studies have noted the important role of market sentiment in affecting asset prices \citep{tetlock_giving_2007, bukovina_social_2016} and market risks \citep{nictoi2020time}. In the most recent years, sentiment analysis has also gained considerable attention with literature increasingly incorporating sentiment information into price forecasting \citep{bollen_twitter_2011, gu_informational_2020}. 

Investor sentiment cannot be directly observed, as it is a latent variable, and various methods have been developed to form the sentiment index in the literature. The first market-based approach proposed by \cite{baker_investor_2006, baker_investor_2007}, who estimated investor sentiment through six market-based measures. The second approach focuses on indices derived from surveys \citep{brown_investor_2004}, such as the surveys published by the American Association of Individual Investors and Investor Intelligence. These two methods have some disadvantages. The first one tends to capture the macroeconomic factors instead of individual investors' attitudes. The second approach usually has a low frequency, as it takes months to collect and analyse the survey data. Based on the Efficient Market Hypothesis (EMH), the market information has probably already been incorporated into asset prices. Therefore, text sentiment analysis has become popular in gauging investor sentiment by extracting text from news and social media platforms \citep{garcia_sentiment_2013}.  Additionally, \cite{da_search_2011} proposed a new measure of investors' attention based on Google search queries. Traditionally, investors access market information through mainstream financial news, official announcements, and analyst reports. However, recent digital developments have enabled social media platforms like Twitter and StockTwits to serve as instant channels for stock information, spreading a larger volume of company-related data to the market at a quicker speed \citep{FT_MissingInnovationMoment_2015, gan_sensitivity_2020}. 

Social media offers a vast, real-time source of public opinion, which is mainly used to facilitate communication and share comments freely with people, compared to the formal news wires/outlets. Twitter (X), for instance, has high popularity in the Western world, where investors, financial analysts, regulators and news agencies share tweets that provide information about the stock market, such as rumours, suggestions and insights \citep{yousaf_quantile_2022}. The information from social media is often based on people's subjective assumptions without solid evidence, rather than verified news, and it is more speculative than news information. These unique characteristics imply that social media may play a different role than news media. Studies have explored the news and social media on the stock market separately at a time, and recently several studies have chosen to compare both media sources together as they believe news and social media have different interactions with asset prices \citep{jiao_social_2020, nyakurukwa_sentimental_2024, nyakurukwa_investor_2025}.

Sentiment can spread among the investors through different platforms \citep{zhang_information_2021}, companies and different markets \citep{baker_global_2012, shiller_narrative_2019, mbarki_sentiment_2022, zhou_stock-level_2023}. Thus, sentiment contagion hypotheses have also been studied as a significant factor affecting the financial market, as there are different channels through which sentiment can spread. We refer to sentiment spillover in this study to represent the sentiment information flow among assets. There are three main streams of the sentiment spillover studied in the literature: sentiment spillover among assets \citep{dash_relationship_2019, plakandaras_spillover_2020, mbarki_sentiment_2022, han_multi-scale_2022}, spillover between sentiment indices and assets \citep{reboredo_impact_2018, song_dynamic_2019,audrino_sentiment_2019, ouyang_measuring_2020, shang_crude_2021, yousaf_quantile_2022, mensi_spillover_2023, jiang_identification_2023, bouteska_contagion_2024, naeem_mapping_2024}, and multilayer networks capturing the interactions among the market return network, volatility network and sentiment network \citep{wang_multilayer_2022, gong_research_2022, gao_investor_2023, li_time-frequency_2024}. 

The sentiment spillover is also measured using different modelling frameworks. In this study, we combine the transfer entropy modelling technique with network theory to quantify the sentiment spillover instead.

The entropy-based method, such as transfer entropy, is widely used in modelling stock price information flow \citep{kwon_information_2008, sandoval_structure_2014, he_comparison_2017}. Later, there are some studies that have adopted entropy-related methods to model the information transfer between assets and sentiment \citep{liu_flow_2020, neto_examining_2022}, but limited studies have considered using entropy-related methods to model sentiment spillover \citep{han_multi-scale_2022}.  The transfer entropy method provides a modelling technique in information flow between random variables that does not assume linearity and normality in the data, and accommodates the information up-to-date. Hence, we believe that the transfer entropy-based modelling technique, together with the network theory, would help us quantify how much sentiment information can transfer from one asset to another, and picture how the information flows within the system.

Previous sentiment spillover literature widely used the network connectedness model proposed by \cite{diebold_measuring_2009, dieboldBetter2012, diebold_network_2014}, along with its subsequent extensions \citep{barunik2018measuring, antonakakis_refined_2020}. Studies also use the network dependency method, which is essentially based on the correlation estimation \citep{wu_dependency_2020, naeem_mapping_2024} and graphical Granger Causality \citep{audrino_sentiment_2019}. However, we argue that the connectedness in the VAR model estimates the variations of the error term. Specifically, it measures how the shocks, which are defined as the standard deviation of the error term in the time series processes, are interconnected through the generalised impulse response function and generalised forecast error variance decomposition estimation on lagged information. In contrast, we aim to measure how sentiment information can be conveyed from one asset to another in a real-time manner. The correlation-based method doesn't allow us to model whether two time series processes have a causal relationship, while Granger Causality can solve this problem, but it requires the assumption of data normality and linearity. Hence, we argue that the property of the transfer entropy method in modelling sentiment spillover is superior to the methods used frequently in the literature.

To the best of our knowledge, there is no such research that addresses and compares the sentiment spillover from both the news sentiment channel and the social media sentiment channel among the companies through a comprehensive transfer entropy network comparison analysis. 

Investor sentiment propagates across companies both directly and indirectly. Companies in the same industry are exposed to similar regulatory environments and customer bases, so policy changes often generate industry-wide sentiment shifts. Meanwhile, news publishers often report several companies together and generalise them towards the industry level \citep{wan_sentiment_2021}. Besides, companies are interconnected through competitive and cooperative relationships \citep{chen_dynamic_2016}, meaning that shocks to one company can signal broader industry conditions. Investors often benchmark one company against their peers. A single company's success or failure can indicate a broader industry's health. Hence, the sentiment between companies interacts in a complex way intrinsically, and we aim to visualise this picture from news and social media channels. 

Therefore, this study aims to answer the following research questions:
\begin{itemize}

    \item How do we measure and quantify the sentiment information spillover between the companies?
        \item How can transfer entropy be integrated with network theory to model sentiment spillover?
        \item How does the time-varying sentiment spillover network pattern evolve over time, and how does it differ between news and social media? 
    
\end{itemize}

Here, we are particularly interested in exploring the sentiment spillover among technology companies in the US during periods of tech hype after the pandemic, such as the growth of Generative AI. We follow \cite{wu_analysing_2024}, who focuses on the volatility contagion among US representative technology companies. We also consider including the rest of the Magnificent Seven companies. Coined by analyst Michael Hartnett from Bank of America in May \citep{Investopedia_Mag7_2025}, the magnificent seven stocks refer to a group of highly influential U.S. companies, particularly from technology sectors. It includes mega-cap tech giants: Alphabet (GOOGL), Amazon (AMZN), Apple (AAPL), Meta Platform (META), Microsoft Corp (MSFT), NVIDIA Corp (NVDA), and Tesla (TSLA). These companies focus on the growth areas such as artificial intelligence, cloud services \citep{Nasdaq_TrustingMag7_2023}, and they have strong market dominant positions \citep{zeng_tail_2025, basele_speculative_2025}.  In this study, we collect the daily News and Twitter sentiment data from the Bloomberg terminal, which provides both news sentiment and social media sentiment indicators for each individual company.

This study has both modelling and empirical contributions to the existing literature. First of all, we introduce a transfer entropy-based network method to build news and social media sentiment spillover networks of US representative technology companies and find that the spillover patterns are different from each other over time. We identify three different spillover regimes from the news sentiment network. We notice a density spike from the end of 2021 to early 2022, whereas this spike doesn't show in the social media sentiment network. Furthermore, we highlight the companies which are the information hub and the maximum possible information flow path in the system. The proposed time-varying sentiment spillover network can be further used to measure the information transfer for different assets to help investors make investment decisions and regulators monitor the density of the information spreading publicly.  

This paper has the following main sections. Section \ref{chap4:definitions} lays out some basic concepts and definitions which will be used throughout this paper. Section \ref{chap4:methodology} introduces the modelling techniques, followed by model calibration. Section \ref{chap4:data} describes the empirical sentiment data used in this study and data preprocessing. Section \ref{results} summarises the research findings based on time-varying sentiment information flow networks and sub-regime analysis to answer our proposed research questions. We conclude the study in Section \ref{conclude}.
\section{Concepts and Definitions}\label{chap4:definitions}
Before we introduce how the sentiment spillover network is constructed, we map out the definitions of network, entropy and transfer entropy, which are the fundamental concepts in constructing the sentiment spillover network. In this section, we consider two time series processes $Y= \bigl\{y_{t}, t=1, 2, \cdots, T\bigr\}$ and $Z=\bigl\{z_{t}, t=1, 2, \cdots, T\bigr\}$ to illustrate. 
\subsection{A network}\label{chap4: network definition}
The directed network is defined as a pair of sets $G=(V, A)$, where $V = \{v_1, v_2, \ldots, v_n\}$, $n\in \mathbb{N}$, is the vertex set and $A = \{(v_i, v_j): v_i, v_j \in V, i\neq j\}$ is the arc set for ordered vertices pairs $(v_i, v_j)$. Let $G=(V,A, w)$ be the arc-weighted, directed network, together with a weighted function $w: A \rightarrow \mathbb{R}^+$. 
This arc-weighted, directed network can be represented by the adjacency matrix $\mathbf{W}$, whose entries we aim to estimate. In this study, the vertex set represents companies, and the weights for the edges represent the sentiment information flow between paired companies, obtained from transfer entropy estimation. 

\subsection{Entropy}
It is necessary to introduce the concept of entropy and conditional entropy before we jump into transfer entropy. Entropy, introduced by \cite{shannon_mathematical_1948}, a pioneer of information theory, is used to measure the average information a random variable contains. This random variable can be treated as an information source. There are various types of information sources across different subject areas, such as a sequence of English text, a sequence of Morse code in telegraph, a stochastic process such as a Markov Chain process, or a time series process. The classical entropy definition is based on the discrete random variables. Therefore, the continuous time series processes are commonly discretised. This is typically done through encoding techniques, such as quantile discretisation. For instance, 
the continuous time process $Z$ can be discretised using quantiles $Q=\lbrace q_k : -\infty < q_1 < q_2 < \cdots < q_K < \infty \rbrace$ that distinguish $K+1$ states. In this study, we set the total number of states to $3$. The probabilities $\Pr(z_t=1, 2, 3)$ of each state (event) occurring are used to calculate the entropy below
\begin{equation}
    Z=z_t=\begin{cases}
1 & \text{for } z_t \leq q_1 \\
2 & \text{for } q_1 < z_t < q_2 \\
3 & \text {for } z_t \geq q_2 
\end{cases}
\end{equation}
\begin{equation}
    H(Z)=-\sum_{z_t\in Z}p(z_t)\log_{2}p(z_t), \hspace{10mm} 0 \leq H(Z) \leq \log_2(|Q|+1),
\end{equation}
where $p(z_t)=P(Z=z_t)$, and the summation is over all possible states of $X$. When $p(z_t=1)=p(z_t=2)=p(z_t=3)=\frac{1}{3}$, entropy $H(Z)$ is the maximum. 

Subsequently, the conditional entropy of $z_{t+1}$ given the previous observations $z_t^{(k)}=  z_t, \allowbreak z_{t-1}, \dots, \allowbreak z_{t-k+1}$ is defined as follows
\begin{equation}
\begin{aligned}
      H(z_{t+1}|z_t^{(k)}) &=-\sum_{z_t\in Z}p(z_{t+1}, z_t^{(k)})\log_2p(z_{t+1}|z_t^{(k)}) \\
      &=-\sum_{z_t\in Z}p(z_{t+1}, z_t^{(k)})\log_2\frac{p(z_{t+1}, z_t^{(k)})}{p(z_t^{(k)})}\\
      &=-\sum_{z_t\in Z}p(z_{t+1}, z_t^{(k)})\log_2p(z_{t+1}, z_t^{(k)})+\sum_{z_t\in Z}p(z_t^{(k)})\log_2p(z_t^{(k)})\\
      &= H(z_{t+1}, z_t^{(k)})- H(z_t^{(k)}), \hspace{10mm} 0\leq H(z_{t+1}|z_t^{(k)})\leq H(Z),
\end{aligned}
\end{equation}
where $p(z_{t+1},z_t^{(k)})$ measures the joint probability of $z_{t+1}$ and $z_t^{(k)}$. The conditional entropy also quantifies how much past information of $z_t^{(k)}$ can be used to predict $z_{t+1}$. We use conditional entropy to calibrate the model by determining the optimal number of previous observations to include. 
\subsection{Transfer Entropy}
We are interested in looking into the information transfer between different sentiment time series, that is, the sentiment spillover.  Therefore, transfer entropy is applied to help us achieve the goal. 
Transfer entropy, proposed by 
\cite{schreiber_measuring_2000} and grounded in information theory, serves as a powerful tool for measuring causal effect. This method is model free, which does not require specific data distribution, such as the normal distribution. Transfer entropy from sentiment time series process $Y$ to $Z$ is defined as follows
\begin{equation}
\begin{aligned}
         T_{Y\rightarrow Z}^{k,l}&=\sum_{z_{t}^{(k)}, \, y_{t}^{(l)}}p(z_{t+1}, z_{t}^{(k)}, y_{t}^{(l)})\log_{2}\frac{p(z_{t+1}|z_{t}^{(k)}, y_{t}^{(l)})}{p(z_{t+1}|z_{t}^{(k)})}\\
         &=\sum_{z_{t}^{(k)},\, y_{t}^{(l)}}p(z_{t+1}, z_{t}^{(k)}, y_{t}^{(l)})\log_{2}p(z_{t+1}|z_t^{(k)}, y_t^{(l)})\\
         &\quad - \sum_{z_{t}^{(k)}, \, y_{t}^{(l)}}p(z_{t+1}, z_{t}^{(k)}, y_{t}^{(l)})\log_{2}p(z_{t+1} | z_t^{(k)})\\
        &=\sum_{z_{t}^{(k)},\, y_{t}^{(l)}}p(z_{t+1}, z_{t}^{(k)}, y_{t}^{(l)})\log_{2}p(z_{t+1}|z_t^{(k)}, y_t^{(l)})\\ &\quad -\sum_{z_{t}^{(k)},\, y_{t}^{(l)}}p(z_{t+1}, z_{t}^{(k)})\log_{2}p(z_{t+1} | z_t^{(k)})\\
        &= H(z_{t+1}|z_t^{(k)})-H (z_{t+1}| z_t^{(k)},\, y_t^{(l)}),  \hspace{10mm} 0 \leq T_{Y\rightarrow Z}^{k,l} \leq H(Z),
\end{aligned}
 \end{equation}
 
This can be interpreted as how much information from lagged time series $y_t^{(l)}= y_t, y_{t-1}, \cdots, y_{t-l+1}$ can be used to predict time series $z_{t+1}$, and $p(z_{t+1},z_{t}^{(k)}, y_{t}^{(l)})$ measures the joint ptobitility of $z_{t+1}$, $z_t^{(k)}$ and $y_t^{(l)}$.  The transfer entropy measures the difference between the conditional entropy $H(z_{t+1}|z_t^{(k)})$ and the conditional entropy of $H (z_{t+1}| z_t^{(k)}, y_t^{(l)})$. 

Here, we normalise the value of transfer entropy to the range $0$ and $1$, which gives a clearer interpretation of the magnitude of information transfer \citep{hmamouche_nlints_2020}. 
The transfer entropy is normalised as follows
\begin{equation}
    \hat{T}_{Y \rightarrow Z}= \frac{T_{Y \rightarrow Z}}{H(z_{t+1}|z_{t})},
\end{equation}
 which indicates the intensity of the information transfer from the sentiment time series $Y$ to the sentiment time series of $Z$. The closer the value is to $1$, the stronger the information flow from time series $Y$ to time series $Z$. It also means that $y_t^{(l)}$ can perfectly predict $z_{t+1}$. 

\section{Methodology}\label{chap4:methodology}

We now outline the construction of the sentiment network based on the estimation and calibration of transfer entropy values. Subsequently, we analyse some properties of the constructed network, such as network density and degree centrality. We finally calibrate our model through quantile discretisation and optimal lag selection.

Let $X_i=\bigl\{x_{i,t}, t=1, 2, \ldots, T, i= 1,2,\ldots, n\bigr\},$ denote the news (or social media) sentiment time series, where each time series process represents the daily sentiment of a specific company $i$. For brevity, we denote $x_{i, t}$ by $x_i$. Therefore, we denote the normalised transfer entropy from sentiment series $x_i$ to $x_j$ as $\hat{T}_{ij}$, where $i=1, 2, \ldots, n,$ and $j = 1, 2, \ldots, n$.  
\subsection{Construct the information flow network}
Our goal is to construct the sentiment information flow network. As the network definition given in Section \ref{chap4: network definition}, we aim to estimate each pair of transfer entropy results and evaluate their significance to obtain the weighted adjacency matrix $\mathbf{W}$: 
\begin{equation}
    \mathbf{W}=
    \begin{bmatrix}
    0 & \hat{T}_{12} & \hat{T}_{13} &\cdots& \hat{T}_{1n} \\
     \hat{T}_{21}  & 0 & \hat{T}_{23} & \cdots& \hat{T}_{2n} \\
    \hat{T}_{31}  & \hat{T}_{32} & 0 & \cdots& \hat{T}_{3n} \\
     \vdots & \vdots & \vdots &\ddots&  \vdots\\
      \hat{T}_{n1}  & \hat{T}_{n2}  & \hat{T}_{n3}  &\cdots& 0\\
    \end{bmatrix}.
\end{equation}

Following \cite{dimpfl_using_2013, behrendt_rtransferentropy_2019}, the significance of the transfer entropy can be assessed using a bootstrap method. The idea is to use the bootstrapped time series of $x_i$ to then calculate the transfer entropy, which can form a distribution. The bootstrapping of $x_i$ is based on simulating Markov processes: the first value is randomly drawn based on the frequency of each encoded symbol in series $x_i$; subsequent values are generated based on the transition probabilities between states. This process continues until the simulated series reaches the same length as the original $x_i$ series. This Markov bootstrap preserves the dependency within $x_i$, but breaks any potential dependencies between $x_i$ and $x_j$. Thus, this allows for testing whether the observed transfer entropy is statistically significant under the null hypothesis that there is no information flow from time series processes $x_i$ to $x_j$. This method also reduces the small sample bias \citep{dimpfl_using_2013}.

Let $N_{\text{boot}}$ denote the total number of bootstrapped transfer entropy values. Let $M$ represent the number of bootstrapped transfer entropy estimates $T_{ij}^{\text{boot}}$ that exceed the estimated $T_{ij}$, based on the specified confidence interval ($1-\alpha$).\footnote{The confidence interval is usually set as $90\%, 95\%$ and $99\%$.} The probability of obtaining a significant transfer entropy value is calculated below: 
 \begin{equation}
     \text{p-value}= \frac{M}{N_{boot}}.
 \end{equation}

If $\text{p-value} < \alpha$, then we reject the null hypothesis and conclude there is information flow from time series $x_i$ to time series $x_j$. This allows us to remove the insignificant values and only keep the significant ones. 


\subsection{Network Analysis}\label{chap4:network indicators}
To understand how information flows within the sentiment networks, we evaluate several fundamental network properties. In particular, we examine network density, the degree distribution, the degree centrality and the maximum spanning arborescence. Together, these measures help us compare how the sentiment spillover network changes over time, which companies consistently act as influential information sources and the dominant sentiment spillover paths.

\subsubsection{Network Density}
 Network density, denoted as $D$, is an important metric to measure how densely the network is connected, and it is bounded between $[0,1]$. A higher density value indicates a more interconnected network where nodes are closely linked, facilitating strong information flow. On the contrary, a lower density value suggests a sparser network with weak information flow. The weighted network density is defined as follows:

\begin{equation}
D=\frac{\sum_{i=1}^n\sum_{j=1}^n\hat{T}_{ij}}{n(n-1)}, \hspace{1cm}0 \leq D \leq 1.
\end{equation}

This can assist us in tracking the density across time windows to compare the aggregated strength of the sentiment information flow, which is useful in identifying the periods of dense information spillover. 

\subsubsection{Weighted in/out-degree}
The weighted in-degree of a vertex $v_i$ is the sum of all the weighted incoming edges, and conversely, the weighted out-degree of a vertex $v_i$ is the sum of all the weighted outgoing edges:
 \begin{equation}
     d_{v_{i}}^{in} = \sum_{j=1}^n \hat{T}_{ji},
 \end{equation}
 \begin{equation}
     d_{v_{i}}^{out} = 
     \sum_{j=1}^n \hat{T}_{ij},
 \end{equation}
Examining the distribution of the weighted in/out-degree of the company helps identify the companies which are consistently information transmitters and absorbers over time. 
\subsubsection{PageRank algorithm}
To further identify influential companies in the sentiment network, we apply the PageRank algorithm \citep{page_pagerank_1999}, which calculates the importance of the vertices in a network. If a vertex is important and also connected to other important vertices, it will have a higher PageRank value. We use this algorithm to select the information hub companies. The PageRank of a vertex $v_i$, is denoted as $PR(v_i)$, is computed as follows on a weighted directed graph:
\begin{equation}
    PR(v_i)= \frac{1-f}{n}+f\sum_{v_j\in V_m}\frac{\hat{T}_{ji}}{d_{v_j}^{\text{out}}}PR(v_j), 
\end{equation}
where $f$ is the damping factor, set as $0.85$ by default, which means there is $0.85$ of chance of a random surfer following an outgoing link on the current page. $n$ is the total number of vertices, $V_m$ is the set of vertices connecting to a vertex $v_i$. $d_{v_j}^{out}$ is weighted out edges from a vertex $v_j$, $\hat{T}_{ji}$ is the weighted edges from $v_j$ to $v_i$. The PageRank algorithm complements the degree measure by identifying how influential the companies are. 

\subsubsection{Maximum Spanning Arborescence Algorithm}
To extract the dominant pathways of the sentiment spillover networks, we adopt the maximum spanning arborescence (MSA) algorithm. It is used to find a spanning arborescence $ \mathbb{T} \subseteq A$ at a root vertex $v_i$ that maximises the total sum of edge weights in the directed network, as proposed by \cite{chu_shortest_1965, jack_optimum_1967}. 

The algorithm starts with finding the root vertex $v_i$ with the highest weighted out-degree edges. Then, only the maximum weighted edge is kept for all other nodes except the root node. It finally returns a directed tree in which each vertex, except the root, has exactly one incoming edge and total weight,  \[\text{max}\sum_{(v_i, v_j) \in \mathbb{T}} w(v_i, v_j),\] is maximised, indicating the strongest information flow path within a network. The final tree not only captures the most influential sentiment spillover pathways, but also highlights the key companies which drive the sentiment spillover with the strongest information flows.

\subsection{Time Varying Transfer Entropy Network Analysis}
Recall that we can obtain a single static information flow network by estimating the paired transfer entropy value between all the companies, which form the edge set $W$. However, a single static network cannot capture the changes in sentiment spillovers over time. 
To answer our research question on how the time-varying sentiment spillover network evolves and differs from news and social media, we need to build dynamic sentiment spillover networks by using a rolling window technique to achieve this. Then, we can build a single network for each time window $w$.  The resulting transfer entropy network matrix under each rolling window $w$, $\mathbf{W}_w$, is a $n \times n$ matrix given by:
\begin{equation}
    \mathbf{W}_w=
    \begin{bmatrix}
    0 & \hat{T}_{12,w} & \hat{T}_{13,w} &\cdots& \hat{T}_{1n,w} \\
     \hat{T}_{21,w}  & 0 & \hat{T}_{23,w} & \cdots& \hat{T}_{2n,w} \\
    \hat{T}_{31,w}  & \hat{T}_{32,w} & 0 & \cdots& \hat{T}_{3n,w} \\
     \vdots & \vdots & \vdots &\ddots&  \vdots\\
      \hat{T}_{n1,w}  & \hat{T}_{n2,w}  & \hat{T}_{n3,w}  &\cdots& 0\\
    \end{bmatrix},
\end{equation}
where each entry $\hat{T}_{ij,w}$ captures the information flow from company $v_i$ to $v_j$ as previously defined in Section \ref{chap4:methodology}. Correspondingly, we can obtain the dynamic network statistics introduced in Section \ref{chap4:network indicators}.
 \subsubsection{Jaccard Similarity Index}\label{jaccard similarity}
 We want to compare further how sentiment contagion networks are different from each other over time. 
 There are different methods proposed in the literature to compare the similarity of networks \citep{tantardini_comparing_2019}. One of the methods is the Jaccard similarity index, developed by Paul Jaccard in 1901 \citep{jaccard1901etude}, which compares the similarity between two sets. 
Suppose there are two networks $G_1=(V, A_1)$ and $G_2=(V, A_2)$, with the same vertex set and different arcs sets $A_1$ and $A_2$ (for simplicity, we don't consider weights here). The Jaccard similarity of two networks is defined as
 
 \begin{equation}
     J(G_1, G_2)=\frac{|A_1 \cap A_2|}{|A_1 \cup A_2| }, \quad J \in [0,1],
 \end{equation}
 where the value equals $1$ represents perfect similarity of two networks. This measure can be used to compare how many edges stay unchanged between two different sentiment networks, suggesting the similarity of the network structures. 

\subsubsection{Summation of power network adjacency matrix}

Another thing we want to check is whether each network shows the final stable structure to identify both direct and indirect information flow links among companies under each rolling window $w$. When the network shows a stable property, which means when all the eigenvalues of network adjacency matrix $\lambda < 1$, we can compute the summation of the adjacency matrix power as the network will converge to a stable point: 
\begin{equation}
\begin{aligned}
\sum_{\omega=1}^\infty \mathbf{W}_w^{\omega} &= \mathbf{W}_w+\mathbf{W}_w^2+\mathbf{W}_w^3+\cdots.\\
\end{aligned}
\label{powermatrix}
\end{equation}

Each entry $\hat{T}_{ij} $ in the adjacency matrix $\mathbf{W}$ captures the edge from vertex $v_i$ to $v_j$. Then, we denote $\mathbf{W}_{ij}^2 $ captures the indirect edges from vertex $v_i$ to $v_j$, such as  $v_i \rightarrow v_k \rightarrow v_j$.
$\sum_{\omega=1}^\infty \mathbf{W}_{ij}^\omega$ calculates the cumulative $\omega$ paths from vertex $v_i$ to vertex $v_j$, which count both direct edge and indirect edges. 

\subsection{Model Calibration}\label{chap4:modelcali}
\subsubsection{Quantile Discretisation and Optimal lag selection}
There are two important parameters that need to be considered carefully when estimating transfer entropy. One is continuous data discretisation, and the other is how to choose optimal lags for two time series processes. 
Transfer entropy is calculated based on the discrete data. Thus, the continuous time series data needs to be discretised. There are different ways of data discretisation, such as the binning method and quantile discretisation.  

If all states $Q$ are equally likely, each state would have a probability of $P(x_t)=\frac{1}{Q}$, which leads to the maximum entropy value. The choice of block lengths or the order of the Markov process $k$ and $l$ in transfer entropy is critical. The idea is to examine how many lags $(x_{t-1},x_{t-2},\ldots,x_{t-k})$ in the time series $x_t$ to include to help predict the present value $x_t$. 
Also, we need to consider that the transition probability is valid, not as 0 in most cases. The sample size matters; if the sample size is small, the number of events would turn out easily to be higher than the total sample size, which leads to $0$ probability for most of the events, making the entropy value meaningless as $0$. We demonstrate this point by simulating an $AR(1)$ process (See the simulation in Appendix \ref{app: AR}.). This process matches the dependence properties of the sentiment series and can guide our choice of lags for the analysis. The simulated sample sizes are 100, 150, 200, 500, 1000, 10000, and 100000. We then use these sample data to calculate their conditional probability against different lags for model calibration. 
\begin{figure}[H]
    \centering
    \includegraphics[width=0.8\linewidth]{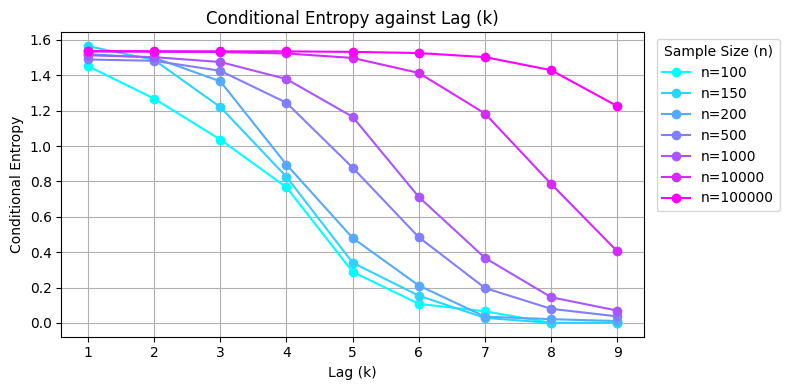}
    \caption{Conditional entropy calibration on different sample sizes against different lags}
    \label{condionalentropy}
\end{figure}
Figure \ref{condionalentropy} plots the number of lags against conditional entropy among different sample sizes. We can notice that the conditional entropy decreased rapidly with the increase of lags for small samples, as the probability of the events tends to become zero. Whereas the conditional entropy stays relatively stable for large samples, it still decreases gradually with the increase of the lag length. However, in reality, it is difficult to get a large sample size. Especially when we estimate time-varying transfer entropy based on the rolling window technique. It is observed that only lag $1$ can be chosen to be the optimal past information used to predict the current period, as further increases in lag result in a drop in conditional entropy. 
\section{Data collection and Pre-processing}\label{chap4:data}
\subsection{Data description}
 Bloomberg Twitter and News sentiment provide a story-level and company-level investor sentiment. Bloomberg News sentiment collects thousands of news from Bloomberg News, web content, and selected premium news wires (more than 50,000 premium online news sources \citep{nyakurukwa_sentimental_2024})\footnote{Premium news wires follow a proprietary model of releasing news. Therefore, Bloomberg cannot further disclose the specific premium news wires (This information is obtained from Bloomberg helpdesk).} for all public-traded companies that get news stories. Bloomberg Twitter sentiment gathers posts from Twitter (Now X) and the Stocktwits (a leading platform for investors and traders) for all public companies which get news stories. We refer to Twitter sentiment as social media sentiment hereafter. To determine the sentiment score of each news article and tweet, Bloomberg trained machine-learning classifiers to assess whether a long-position investor would interpret the content as bullish, bearish or neutral. Each tweet and news article is assigned a sentiment polarity score and confidence. The score is a categorical value ($S \in \{-1, 0, 1\}$). The confidence value ($C$) ranges from 0 to 100, representing the likelihood of the sentiment classification. These story-level sentiments are then aggregated by a confidence-weighted average of the individual scores between days $t-1$ and $t$, i.e., within $(t-1,t]$. This produces a daily sentiment score for each company as follows (see \citep{cui2016embedded, gu_informational_2020, garcia_alvarado_detecting_2022}, for more details)
\begin{equation}
    \text{Sentiment}_{i,t}=\frac{\sum_{k\in K(i,T)} S_i^kC_i^k}{K_{i,T}}, \hspace{10mm} T\in (t-1, t],
\end{equation}
where $K_{i,T}$ is company $i$'s total number of news or posts during time $T$.

We collect news and social media sentiment for selected technology companies from the Bloomberg Terminal. Following \cite{wu_analysing_2024}, we collect the corresponding news and social Media sentiment data of the corresponding technology companies from the Bloomberg terminal, along with the remaining Magnificent Seven companies. The list of 34 companies can be found in the Appendix \ref{appen: company lists}. The sentiment time series spans from September 30, 2019 to October 17, 2024, totalling 1319 sentiment data points. 

As addressed before, the sentiment score varies from $[-1,1]$, providing a numerical continuous indicator of investor sentiment, where $-1$ reflects the investor's attitude towards a company having a strongly negative sentiment, $0$ represents neutral sentiment, and $1$ indicates strongly positive sentiment. 

\subsection{Missing data imputation}
Since there is not always news or social media activity for every company each day, Bloomberg may occasionally be unable to capture news or social media feed for a given company on a particular day. In such cases, the sentiment score for that day is recorded as a missing value. Therefore, some of our collected news and social media sentiment time series show a varying proportion of missing data, ranging from $0.758\%$ to $24.56\%$ across different companies.

There are different ways to interpolate or impute missing data, depending on the nature of the data and the underlying reasons for the missingness. In our case, we consider that investors' reactions towards news are non-linear, which tend to decay over time. Specifically, sentiment tends to decrease non-linearly over time, which means the initial responses are strong, but gradually fade. As a result, we assume the sentiment exhibits an exponential decay over time. Accordingly, we propose the following decay-based imputation function, which estimates the missing sentiment values using the most recently available observation: 
    \begin{equation}
        x_{t} = Be^{-\lambda t} + \epsilon_t, \hspace{1cm} \epsilon_t \sim \mathcal{N}(0,\sigma^2),
    \end{equation}
   where $B$ denotes the most recent available non-missing value prior to time $t$, $\lambda$ controls the decay rate, set to 0.23 based on half-life calibration. We compare two ways of missing data imputation. One is decay without noise, so the error term is set to 0, the other one is decay including a random noise term in the function to reduce the artificiality of the final imputed data. The standard deviation $\sigma$ of the error term is estimated from its original time series by fitting an AR model (see Appendix \ref{app: AR} for details). 
   \begin{figure}[H]
       \centering
    \begin{subfigure}{0.48\textwidth}
           
       \includegraphics[width=\linewidth]{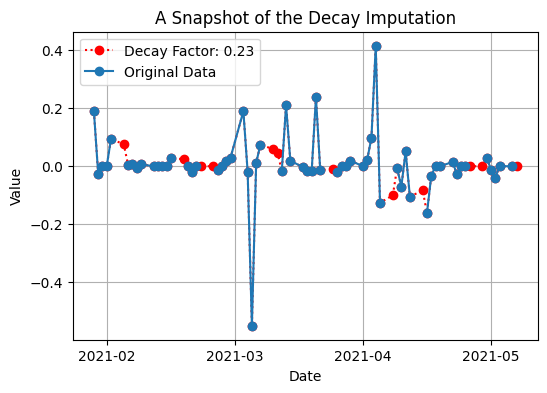}
       \caption{Decay imputation}
       \label{decayimpute}
   \end{subfigure}
\hfill
   \begin{subfigure}{0.48\textwidth}
       \centering

    \includegraphics[width=\linewidth]{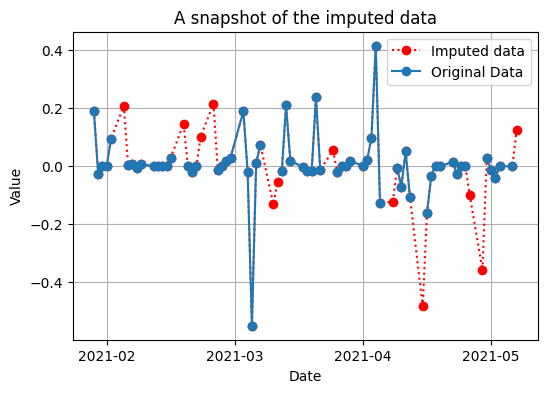}
       \caption{Decay imputation with random noise}
       \label{imputeddata}
    \end{subfigure}
    \caption{A comparison of missing data decay imputation and decay imputation with random noise.}\caption*{\small Notes: This figure visualises an example segment of the missing data decay imputation and decay imputation with random noise on MCO company from 2021-02-20 to 2021-06-01.}
   \end{figure}

Figure \ref{decayimpute} and Figure \ref{imputeddata} compare an example zoomed-in social media sentiment series segment for Moody's Corp (MCO). From the plots, we can see that the original sentiment series does not always show a decay trend. Therefore, it is more rational to add a random noise term during the imputation. The following analysis is based on the imputed data, with the addition of a random noise term. 

\subsection{Sentiment data properties}
\begin{figure}[H]
   \centering
     \includegraphics[width=\linewidth]{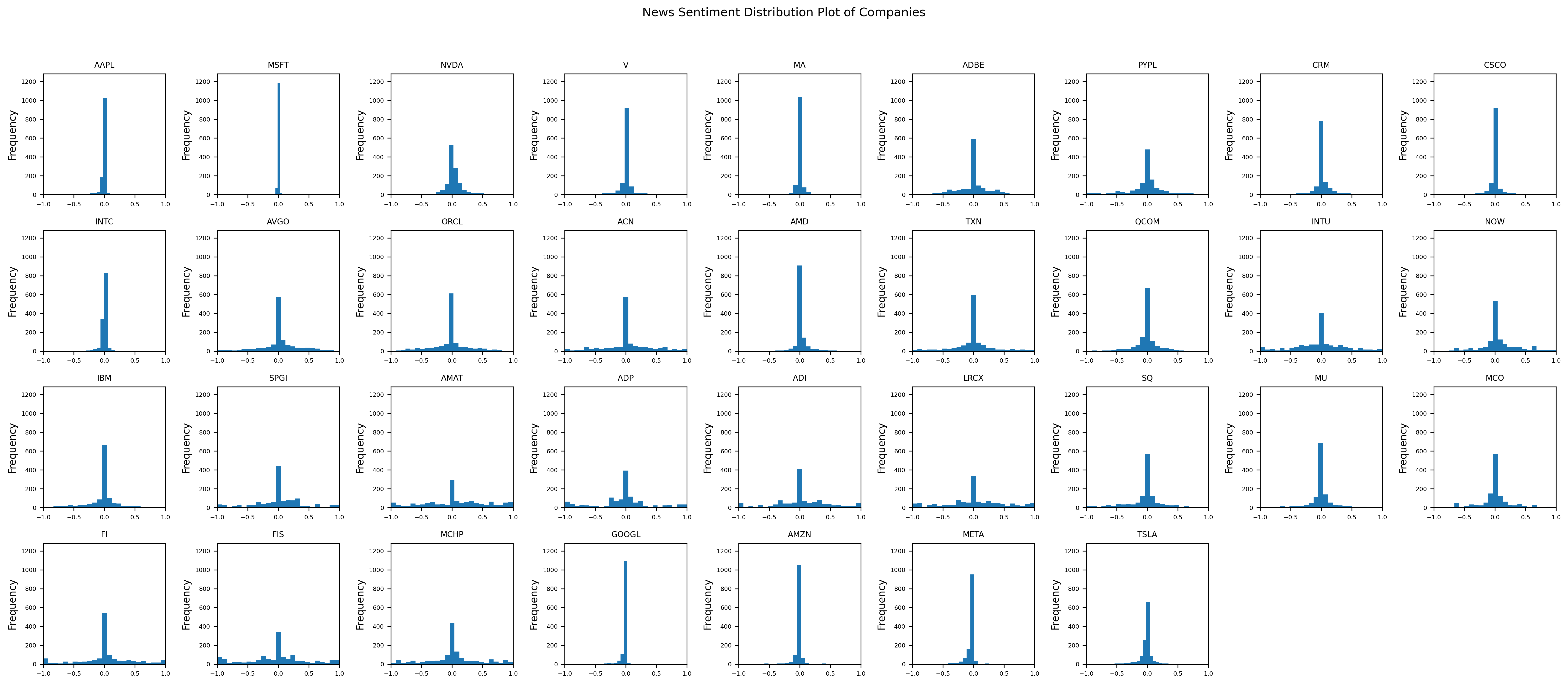}
    \caption{Distribution plot of News sentiment on each company}
    \label{fig:news_distribution}
\end{figure}

\begin{figure}[H]
    \centering
     \includegraphics[width=\linewidth]{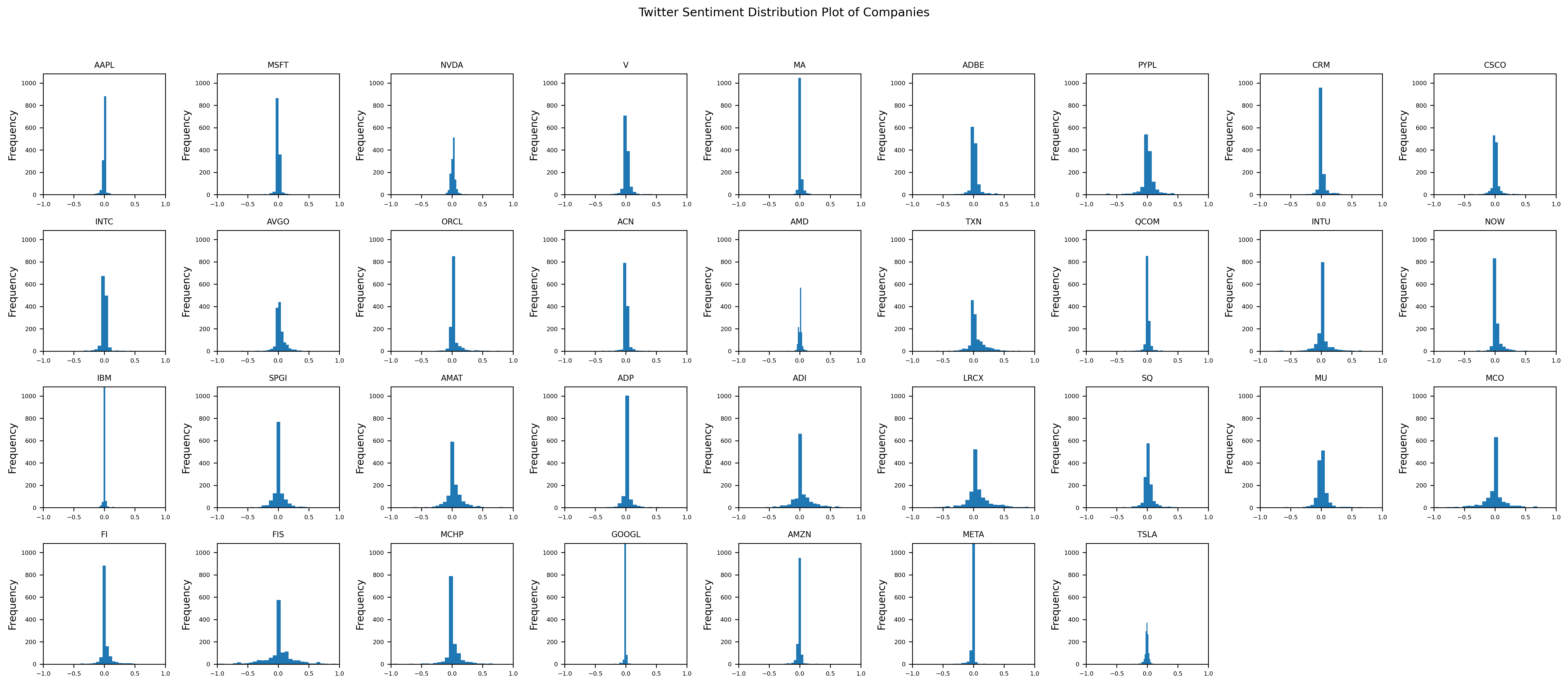} 
    \caption{Distribution plot of Social Media sentiment on each company}
    \label{fig:twitter_distribution}
\end{figure}

Below, we provide distribution plots for news sentiment (Figure \ref{fig:news_distribution}) and social media sentiment (Figure \ref{fig:twitter_distribution}). We can first see that most of the sentiment data show non-normal, left or right-skewed and heavily-tailed, suggesting non-Gaussian behaviour.
 
The Augmented Dickey-Fuller (ADF) tests in the summary statistics tables (Table \ref{tab:newsdes} and \ref{tab:twides}) show that all the sentiment data are stationary. The Shapiro-Wilk tests on sentiment time series also reject the null hypothesis of normal distribution of the data. To take these together, the empirical properties of the data therefore confirm our modelling choice to analyse sentiment spillover. 
\begin{table}[H]
\centering
\caption{Descriptive statistics of News sentiment for the companies.}
\begin{adjustbox}{max width=\textwidth}
\Large
\begin{tabular}{lrrrrrcccccc}
\toprule
\textbf{Ticker} & \textbf{Min} & \textbf{Max} & \textbf{Mean} & \textbf{SD} & \textbf{Q25} & \textbf{Q75} & \textbf{Skewness} & \textbf{Kurtosis} & \textbf{ADF stat} & \textbf{Shapiro stat} \\
\midrule
AAPL & -0.875 & 0.466 & -0.0118 & 0.0665 & -0.0113 & 0.0017 & -3.559 & 42.952 & $-26.420^{***}$ & $0.468^{***}$ \\
MSFT & -0.287 & 0.581 & 0.00224 & 0.0425 & -0.0014 & 0.0011 & 5.034 & 61.567 & $-29.2186^{***}$ & $0.302^{***}$ \\
NVDA & -0.839 & 0.957 & 0.0369 & 0.181 & -0.0149 & 0.0766 & 0.844 & 6.440 & $-10.563^{***}$ & $0.823^{***}$ \\
V    & -0.992 & 0.876 & -0.000929 & 0.137 & -0.00675 & 0.0133 & -0.690 & 14.431 & $-11.833^{***}$ & $0.644^{***}$ \\
MA   & -0.814 & 0.962 & 0.00462 & 0.0934 & -0.0051 & 0.00655 & 2.298 & 36.287 & $-17.644^{***}$ & $0.519^{***}$ \\
ADBE & -0.986 & 0.982 & -0.00676 & 0.279 & -0.0742 & 0.068 & -0.185 & 1.846 & $-5.945^{***}$ & $0.912^{***}$ \\
PYPL & -1.000 & 0.993 & -0.0307 & 0.318 & -0.105 & 0.0750 & -0.420 & 1.850 & $-6.210^{***}$ & $0.900^{***}$ \\
CRM  & -0.853 & 1.000 & 0.0275 & 0.186 & -0.009 & 0.0378 & 0.967 & 6.809 & $-29.957^{***}$ & $0.751^{***}$ \\
CSCO & -0.917 & 0.940 & 0.00295 & 0.167 & -0.00915 & 0.0102 & 0.177 & 11.050 & $-9.876^{***}$ & $0.641^{***}$ \\
INTC & -0.656 & 0.825 & -0.0106 & 0.0817 & -0.0092 & 0.0026 & -0.560 & 33.103 & $-21.670^{***}$ & $0.504^{***}$ \\
AVGO & -1.000 & 0.999 & 0.0350 & 0.311 & -0.0213 & 0.113 & -0.046 & 1.820 & $-33.855^{***}$ & $0.898^{***}$ \\
ORCL & -0.999 & 0.970 & -0.0211 & 0.295 & -0.0955 & 0.0276 & -0.112 & 1.416 & $-25.609^{***}$ & $0.900^{***}$ \\
ACN  & -1.000 & 1.000 & 0.0172 & 0.365 & -0.0508 & 0.123 & 0.003 & 0.939 & $-33.0278^{***}$ & $0.916^{***}$ \\
AMD  & -0.829 & 0.968 & 0.0254 & 0.137 & -0.00405 & 0.0266 & 1.652 & 12.148 & $-11.821^{***}$ & $0.662^{***}$ \\
TXN  & -0.999 & 1.000 & -0.014  & 0.324 & -0.0789 & 0.0505 & -0.114 & 2.02 & $-19.700^{***}$ & $0.877^{***}$ \\
QCOM & -0.98 & 0.99 & -0.0129 & 0.213 & -0.0468 & 0.0249 & -0.274 & 5.05 & $-11.400^{***}$ & $0.828^{***}$ \\
INTU & -1.000 & 1.000 & -0.023 & 0.402 & -0.22 & 0.164 & -0.084 & 0.529 & $-32.000^{***}$ & $0.959^{***}$ \\
NOW  & -0.993 & 1.000 & 0.0612 & 0.308 & -0.0244 & 0.164 & 0.317 & 1.43 & $-13.000^{***}$ & $0.907^{***}$ \\
IBM  & -0.992 & 0.989 & -0.0308 & 0.287 & -0.059 & 0.0245 & -0.329 & 2.81 & $-6.790^{***}$ & $0.846^{***}$ \\
SPGI & -1.000 & 1.000 & 0.00737 & 0.392 & -0.136 & 0.206 & -0.163 & 0.798 & $-12.200^{***}$ & $0.941^{***}$ \\
AMAT & -1.000 & 1.000 & 0.0592 & 0.502 & -0.277 & 0.397 & -0.11 & -0.503 & $-15.600^{***}$ & $0.972^{***}$ \\
ADP  & -0.998 & 0.998 & -0.0394 & 0.433 & -0.204 & 0.108 & -0.091 & 0.54 & $-31.000^{***}$ & $0.923^{***}$ \\
ADI  & -1.000 & 1.000 & 0.0241 & 0.425 & -0.162 & 0.26 & -0.08 & 0.384 & $-7.190^{***}$ & $0.954^{***}$ \\
LRCX & -1.000 & 1.000 & 0.0139 & 0.473 & -0.212 & 0.277 & -0.054 & -0.181 & $-6.890^{***}$ & $0.963^{***}$ \\
SQ   & -0.991 & 1.000 & -0.0299 & 0.291 & -0.0846 & 0.052 & -0.396 & 2.34 & $-4.420^{***}$ & $0.883^{***}$ \\
MU   & -0.993 & 0.982 & -0.00516 & 0.249 & -0.0329 & 0.0346 & -0.247 & 4.61 & $-14.500^{***}$ & $0.800^{***}$ \\
MCO  & -0.982 & 1.000 & 0.00232 & 0.282 & -0.0549 & 0.0693 & -0.002 & 2.3 & $-8.090^{***}$ & $0.882^{***}$ \\
FI   & -1.000 & 1.000 & 0.0211 & 0.408 & -0.048 & 0.147 & -0.167 & 0.967 & $-20.800^{***}$ & $0.904^{***}$ \\
FIS  & -1.000 & 1.000 & -0.0258 & 0.473 & -0.23 & 0.204 & -0.106 & -0.023 & $-5.000^{***}$ & $0.951^{***}$ \\
MCHP & -1.000 & 1.000 & 0.0194 & 0.403 & -0.107 & 0.146 & -0.004 & 0.540 & $-29.100^{***}$ & $0.933^{***}$ \\
GOOGL & -0.547 & 0.277 & -0.00606 & 0.034 & -0.0054 & -0.001 & -7.14 & 116 & $-15.200^{***}$ & $0.313^{***}$ \\
AMZN & -0.800 & 0.371 & -0.00935 & 0.0611 & -0.015 & 0.00815 & -5.250 & 62.100 & $-6.500^{***}$ & $0.482^{***}$ \\
META & -0.862 & 0.465 & -0.0154 & 0.0702 & -0.0111 & -0.0009 & -6.78 & 71.2 & $-6.210^{***}$ & $0.323^{***}$ \\
TSLA & -0.321 & 0.164 & -0.0105 & 0.0378 & -0.0209 & 0.00915 & -2.11 & 14.5 & $-7.310^{***}$ & $0.817^{***}$ \\
\bottomrule
\end{tabular}
\end{adjustbox}

\begin{minipage}{\linewidth}
\justifying
\small
Notes: This table reports the summary statistics for the news sentiment data, including minimum, maximum, mean, standard deviation, $25\%$ and $75\%$ quantiles, skewness and kurtosis. We also report the Augmented Dickey-Fuller test statistics and Shapiro-test statistics. $^{***}$ indicate the significance level at $1\%$. 
\end{minipage}
\label{tab:newsdes}
\end{table}

\begin{table}[H]
\centering
\caption{Descriptive statistics of Social Media sentiments for the companies.}
\begin{adjustbox}{max width=\textwidth}
\Large
\begin{tabular}{lrrrrrcccccc}
\toprule
\textbf{Ticker} & \textbf{Min} & \textbf{Max} & \textbf{Mean} & \textbf{SD} & \textbf{Q25} & \textbf{Q75} & \textbf{Skewness} & \textbf{Kurtosis} & \textbf{ADF stat} & \textbf{Shapiro stat} \\
\midrule
AAPL  & -0.687 & 0.387 & -0.00659 & 0.0446 & -0.00880 & 0.00170 & -3.80 & 66.5 & $-4.97^{***}$ & $0.489^{***}$ \\
MSFT  & -0.829 & 0.644 & 0.000479 & 0.0511 & -0.00540 & 0.00700 & -3.58 & 106 & $-8.07^{***}$ & $0.414^{***}$ \\
NVDA  & -0.403 & 0.432 & 0.0174 & 0.0469 & -0.00560 & 0.0336 & 0.664 & 15.2 & $-6.89^{***}$ & $0.840^{***}$ \\
V     & -0.870 & 0.689 & 0.0118 & 0.0830 & -0.00400 & 0.0257 & -1.15 & 33.9 & $-28.4^{***}$ & $0.605^{***}$ \\
MA    & -0.444 & 0.823 & 0.0121 & 0.0634 & -0.000500 & 0.0110 & 5.09 & 60.6 & $-28.5^{***}$ & $0.408^{***}$ \\
ADBE  & -0.820 & 0.837 & 0.0185 & 0.0896 & -0.00250 & 0.0291 & 1.01 & 23.5 & $-10.4^{***}$ & $0.650^{***}$ \\
PYPL  & -0.865 & 1.01 & 0.0170 & 0.140 & -0.0122 & 0.0435 & 0.472 & 12.6 & $-17.6^{***}$ & $0.744^{***}$ \\
CRM   & -0.929 & 0.742 & 0.00868 & 0.0820 & -0.00120 & 0.0139 & -1.74 & 37.6 & $-16.3^{***}$ & $0.540^{***}$ \\
CSCO  & -0.551 & 0.640 & 0.00649 & 0.0898 & -0.00520 & 0.0191 & 0.0477 & 14.4 & $-7.29^{***}$ & $0.672^{***}$ \\
INTC  & -0.970 & 0.751 & -0.00187 & 0.0776 & -0.0117 & 0.0139 & -2.07 & 46.7 & $-9.95^{***}$ & $0.532^{***}$ \\
AVGO  & -0.600 & 0.691 & 0.0315 & 0.100 & 0.00 & 0.0575 & 0.224 & 10.4 & $-16.4^{***}$ & $0.783^{***}$ \\
ORCL  & -0.779 & 0.777 & 0.0261 & 0.102 & 0.00 & 0.0245 & 2.00 & 19.6 & $-7.35^{***}$ & $0.573^{***}$ \\
ACN   & -0.846 & 0.658 & 0.00677 & 0.0718 & 0.00 & 0.0117 & -1.32 & 45.6 & $-21.7^{***}$ & $0.442^{***}$ \\
AMD   & -0.188 & 0.338 & 0.00889 & 0.0354 & -0.0111 & 0.0203 & 1.58 & 17.1 & $-7.83^{***}$ & $0.802^{***}$ \\
TXN   & -0.616 & 0.816 & 0.0491 & 0.139 & 0.00 & 0.0811 & 1.16 & 6.06 & $-16.5^{***}$ & $0.818^{***}$ \\
QCOM  & -0.621 & 0.568 & 0.00425 & 0.0703 & -0.00300 & 0.0153 & -1.40 & 30.2 & $-20.7^{***}$ & $0.537^{***}$ \\
INTU  & -0.845 & 0.835 & 0.00963 & 0.130 & -0.00210 & 0.0182 & -0.367 & 12.4 & $-13.3^{***}$ & $0.691^{***}$ \\
NOW   & -0.665 & 0.912 & 0.0269 & 0.0961 & 0.00 & 0.0292 & 2.86 & 27.4 & $-7.36^{***}$ & $0.580^{***}$ \\
IBM   & -0.441 & 0.413 & 0.00241 & 0.0359 & -0.00165 & 0.00470 & 2.11 & 68.2 & $-21.9^{***}$ & $0.380^{***}$ \\
SPGI  & -0.950 & 0.895 & 0.00475 & 0.131 & -0.0141 & 0.0249 & -0.920 & 15.1 & $-11.7^{***}$ & $0.717^{***}$ \\
AMAT  & -0.884 & 0.952 & 0.0351 & 0.161 & -0.00465 & 0.0744 & 0.685 & 8.54 & $-32.4^{***}$ & $0.809^{***}$ \\
ADP   & -0.950 & 0.932 & 0.0129 & 0.0950 & 0.00 & 0.00795 & 1.78 & 29.5 & $-20.9^{***}$ & $0.534^{***}$ \\
ADI   & -0.865 & 0.941 & 0.0342 & 0.179 & 0.00 & 0.0708 & 0.819 & 5.50 & $-14.4^{***}$ & $0.827^{***}$ \\
LRCX  & -0.908 & 1.03 & 0.0643 & 0.212 & 0.00 & 0.119 & 0.673 & 4.27 & $-29.4^{***}$ & $0.864^{***}$ \\
SQ    & -0.655 & 0.831 & 0.0233 & 0.110 & -0.0162 & 0.0418 & 1.47 & 12.5 & $-5.04^{***}$ & $0.753^{***}$ \\
MU    & -0.961 & 0.842 & 0.0145 & 0.115 & -0.0219 & 0.0425 & 0.191 & 15.5 & $-6.70^{***}$ & $0.752^{***}$ \\
MCO   & -0.982 & 0.950 & -0.00850 & 0.198 & -0.0391 & 0.0198 & -0.194 & 5.47 & $-30.1^{***}$ & $0.826^{***}$ \\
FI    & -0.706 & 0.855 & 0.0196 & 0.102 & 0.00 & 0.0212 & 1.25 & 17.1 & $-21.6^{***}$ & $0.633^{***}$ \\
FIS   & -0.999 & 0.950 & 0.00960 & 0.241 & -0.0254 & 0.0914 & -0.267 & 3.24 & $-14.6^{***}$ & $0.881^{***}$ \\
MCHP  & -0.964 & 0.993 & 0.0187 & 0.151 & 0.00 & 0.0258 & 0.0372 & 13.5 & $-11.5^{***}$ & $0.668^{***}$ \\
GOOGL & -0.945 & 0.396 & -0.0216 & 0.0795 & -0.0141 & -0.0008 & -4.473 & 38.501 & $-17.063^{***}$ & $0.416^{***}$ \\
AMZN  & -0.780 & 0.894 & -0.0117 & 0.101 & -0.0145 & 0.003 & -0.255 & 24.998 & $-11.127^{***}$ & $0.517^{***}$ \\
META  & -0.964 & 0.556 & -0.0474 & 0.123 & -0.0488 & -0.0038 & -3.180 & 19.087 & $-10.120^{***}$ & $0.552^{***}$ \\
TSLA  & -0.796 & 0.503 & -0.0266 & 0.131 & -0.0375 & 0.0088 & 1.551 & 8.537 & $-6.145^{***}$ & $0.748^{***}$ \\
\bottomrule
\end{tabular}
\end{adjustbox}

\begin{minipage}{\linewidth}
\justifying
\small
Notes: This table reports the summary statistics for the social media sentiment data, such as minimum, maximum, mean, standard deviation, $25\%$ and $75\%$ quantile, skewness and kurtosis. We also report the Augmented Dickey-Fuller test statistics and Shapiro-test statistics. $^{***}$ indicate the significance level at $1\%$. 
\end{minipage}
\label{tab:twides}
\end{table}

\section{Results}\label{results}
In this section, we present the findings for the dynamic news and social media sentiment information flow networks based on the rolling window analysis. We compare the network density over time and highlight the nodes that consistently serve as information hubs. By analysing changes in network density, we identify three different regimes. Consequently, we conduct a comprehensive regime-specific analysis of both news and social media networks, including degree centrality measures and a Maximum Spanning Arborescence analysis.

\subsection{Dynamic sentiment information flow network}\label{dsif}

\begin{figure}[H]
\begin{subfigure}{0.47\textwidth}
        \centering
    \includegraphics[width=\textwidth]{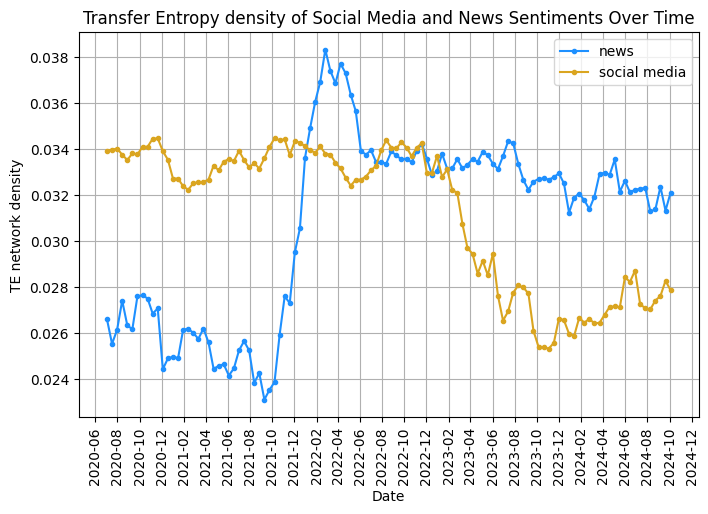}
  \caption{Without filtering the insignificant values}
\end{subfigure}
\hfill
\begin{subfigure}{0.50\textwidth}
    \centering
\includegraphics[width=\textwidth]{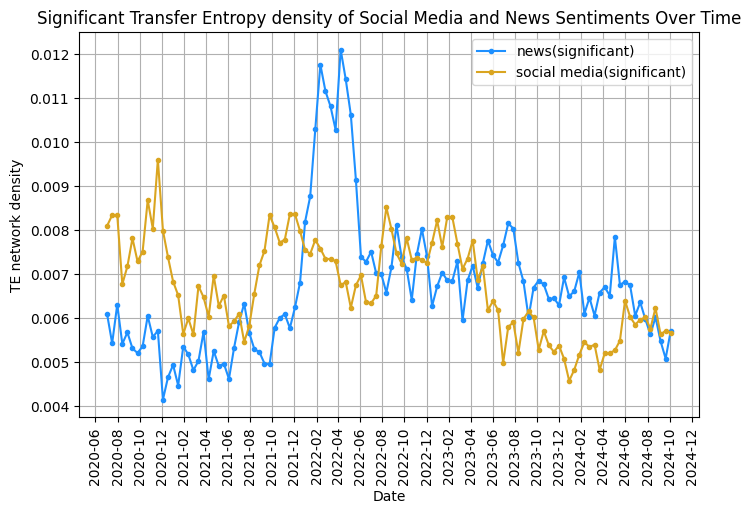}
\caption{Only keep the significant values}
    \end{subfigure}
    \caption{Transfer entropy network density over time.}\caption*{\small Notes: The plot on the left-hand side shows the density of fully connected time-varying sentiment transfer entropy networks. The blue and yellow lines indicate news sentiment networks and social media sentiment networks, respectively. Each point represents the cumulative transfer entropy network constructed using data from the past 200 days. The plot on the right-hand side represents the same concept; however, we only keep the statistically significant transfer entropy values at $10\%$ significance level in the network, and then the density is calculated based on the filtered networks accordingly. }
    \label{te_sig_density}
\end{figure}

We focus on 34 technology companies as mentioned in the section \ref{chap4:data}. Transfer entropy is calculated in pairs, so there are a total of 1122 transfer entropy values as sentiment spillover network entries. We aim to build dynamic sentiment networks based on news and social media separately to observe how the information is transferred over time and compare whether the information flow in news and social media follows the same trend. Thus, a rolling window technique is adopted. We set the rolling window length as 200 days, 40 trading weeks, roughly a trading year, with a step size of 10 days. It is crucial to include sufficient data observations to obtain valid transfer entropy estimates. Based on our model calibration, we believe 200 days is a proper rolling window to choose from both theoretical evidence and financial intuition according to the literature \citep{audrino_sentiment_2019}. Figure \ref{te_sig_density} shows the network density of information flow in news and social media, which overlooks the pattern of information transfer of these two different media sources. At any given time $t$, the information flow captures the cumulative effects of the past 200 days. 

We notice a sharp increase in the network density of the news information transfer among technology companies after October 2021.  Although it bounces back slightly around April 2022, the density then remains at a relatively higher level compared to the period before the climbing period. 

However, the social media network density shows a different trend. After October 2022, there is an incredible decrease in social media sentiment density. The network density is higher than the news network density before December 2021, and it intertwines with the news network density between August 2022 and February 2023. After this period, it falls below the density observed in the news network. We filter out the noise by only keeping significant transfer entropy values, and the confidence interval is set as 90\%. On the right-hand side of Figure \ref{te_sig_density} shows the news and social media information transfer network density is shown based on the significant transfer entropy values. The overall density network after filtering out the significant value is very low, referring to a very sparse sentiment information network. However, the significant upward shift in the news sentiment network remains evident, and there is still a slight downward shift in the social media sentiment network.

\begin{figure}[H]
    \centering
    \begin{subfigure}{0.42\linewidth}
        \centering
        \includegraphics[width=\textwidth]{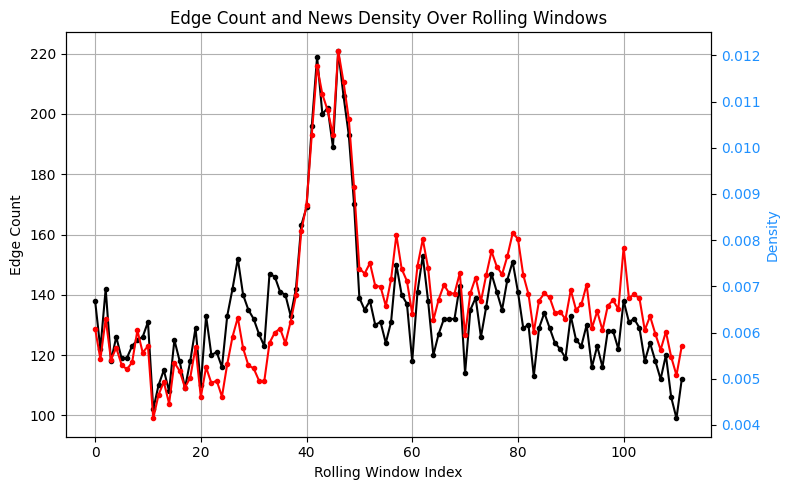}
        \caption{News}
        \label{edges_n}
    \end{subfigure}
    \hfill
    \begin{subfigure}{0.42\linewidth}
        \centering
        \includegraphics[width=\textwidth]{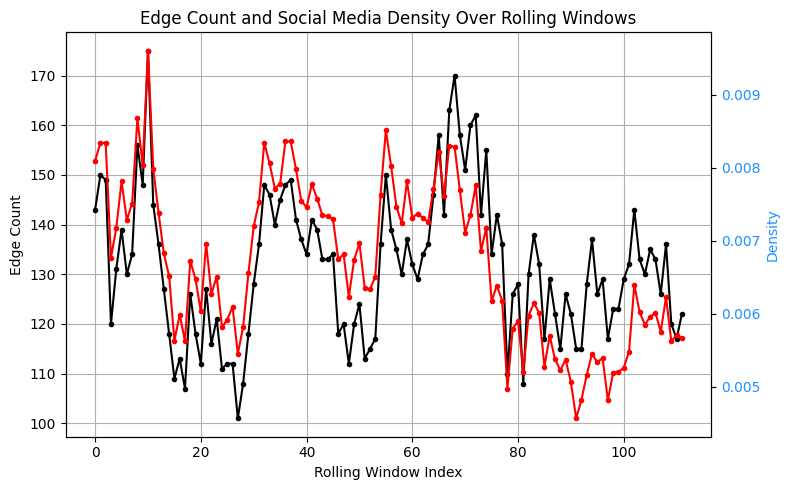}
        \caption{Social Media}
        \label{edges_t}
    \end{subfigure}
    \caption{Edges count and density of news and social media transfer entropy network over time.}\caption*{\small Notes: In both plots, the red lines represent network density, while the black lines indicate the number of edges over time.}
    \label{fig:edge_counts_density}
\end{figure}

\begin{figure}[H]
    \centering
    \includegraphics[width=0.8\linewidth]{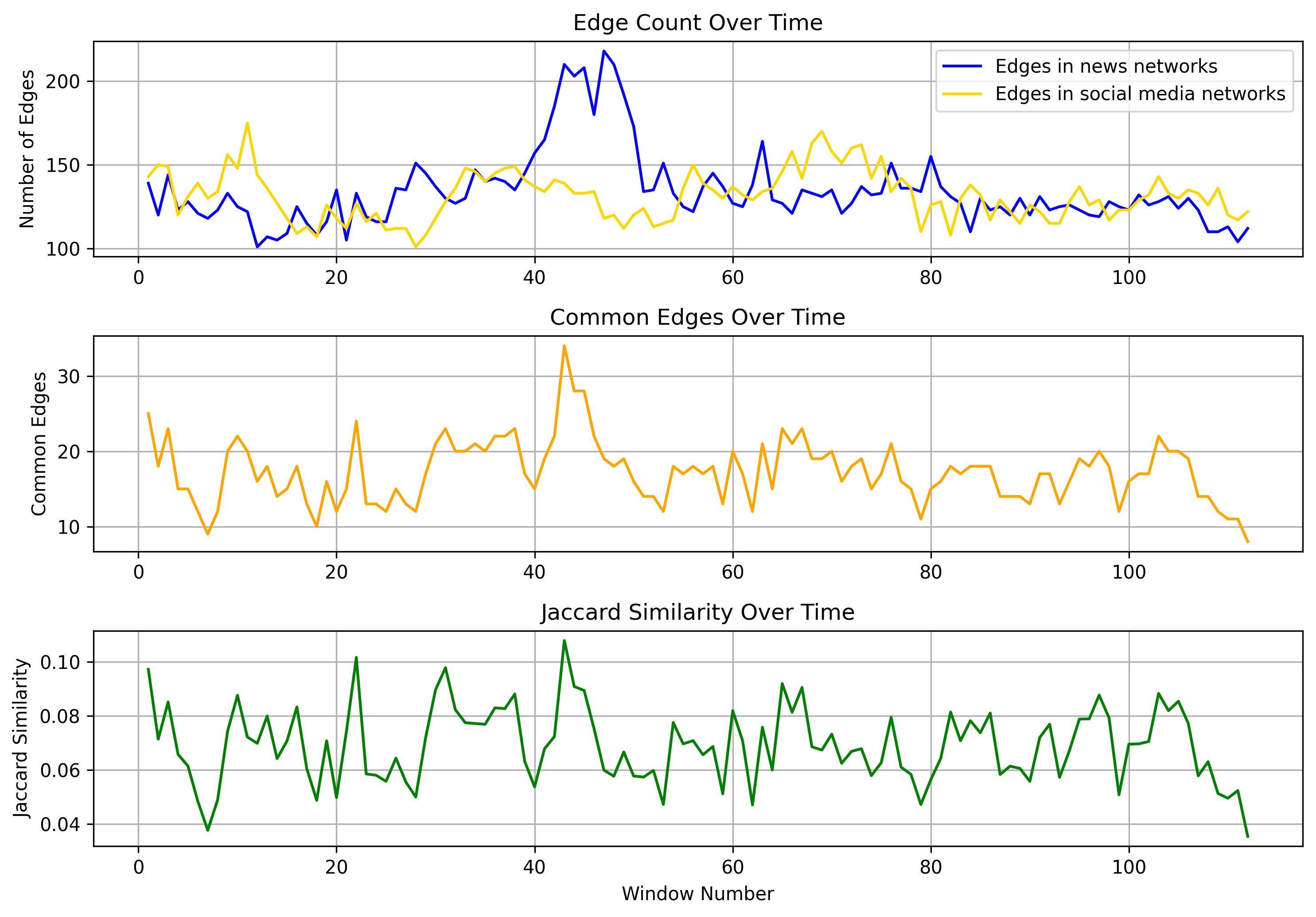}
    \caption{Jaccard Similarity scores over time between two different media sources.}
    \caption*{\small Notes: This figures reports the total number of edges in the sentiment spillover network over time, the common edges between the news media and social media under each window and the Jaccard similarity scores over time. }
    \label{fig:jaccard similarity}
\end{figure}

Figure \ref{fig:edge_counts_density} shows that the number of edges in the network follows a similar pattern to the density plot. We later compared whether the news and social media sentiment networks share some common edges within each window using the Jaccard similarity measure (see section \ref{jaccard similarity}). 
The result presented in Figure \ref{fig:jaccard similarity} indicates that they do not share a common network structure with a very low Jaccard similarity score, which further supports the fact that news and social media originate from two different root information sources. 

\begin{figure}[H]
    \centering
    \includegraphics[width=\linewidth]{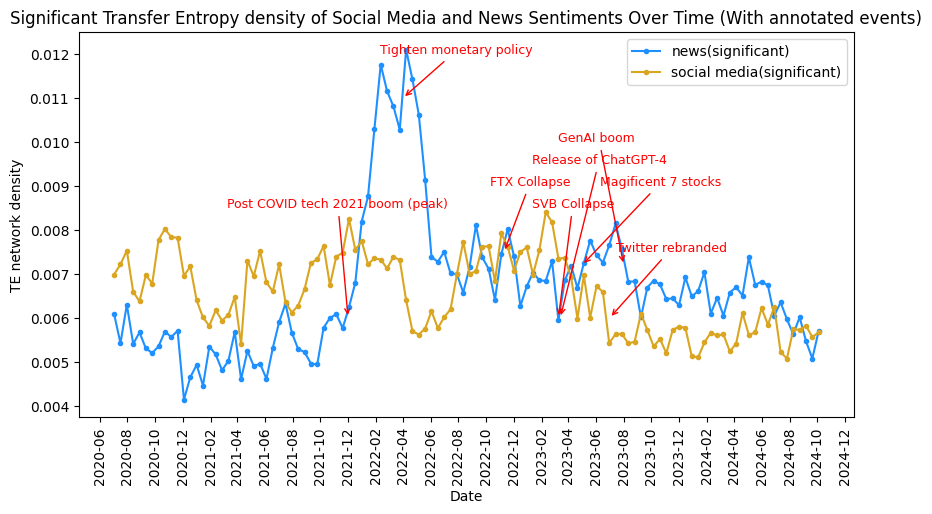}
    \caption{Sentiment network plot with annotated events.} \caption*{\small Notes: This figure highlights several major events that may have influenced the sentiment spillover among the technology companies. In the aftermath of COVID, the technology sector entered a period of pronounced hype toward the end of 2021. Starting in 2022, the U.S. Federal Reserve announced the tightening of monetary policy. In November 2022, FTX announced its collapse. Subsequently, SVB announced its failure in March 2023. During the same time period, starting in March 2023, generative AI hype gained significant momentum with the release of ChatGPT-4. And later in May 2023, Bank of America coined the \enquote{Magnificent Seven} to acknowledge its strong market position of seven leading technology firms.
    In addition, Twitter rebranded to X in July 2023.}
    \label{annotated events}
\end{figure}

We then explore the potential factors that explain the observed pattern in the significant network density plot over time. Figure \ref{annotated events} displays several major events that coincide with fluctuations in the density of sentiment information flow among the tech companies. 

The news sentiment spillover spike peaked from February 2022 to May, which reflects the aggregated information transmission over the past 200 days, as addressed before. After the COVID-19 pandemic period, the technology sector had shown spectacular performance until the end of 2021 \citep{CPRAM_TechCycle_2024}. COVID-19 pandemic caused another history of human civilisation, people had to change their behaviour instantly and adapt to a different reality, powered by tech \citep{McKinsey_DigitalEdge_2021, hossain_covid-19_2023, lee_financial_2023}. 

The market capitalisation of the tech industry, the AI market size worldwide and the revenue of the Magnificent Seven company have increased significantly in 2021. Therefore, it is reasonable to think that the overall tech industry and tech companies would appear more in the news, and the news agencies tend to report several tech companies together during that period, considering their competition or cooperation within the same industry \citep{wan_sentiment_2021}. However, this wave of hype didn't last long, as we can also observe from the news density line, which drops after. Starting from March 2022, the US Federal Reserve decided to tighten the monetary policy to tackle the high inflation problem, which hit the overall technology industry (see \citep{CPRAM_TechCycle_2024}). This monetary policy affects all industries on a macro level.  However, if we turn our spotlight to social media during this period. The sentiment spillover pattern is distinct from what we can observe in news spillover, whereas it shows a slight decrease and then returns. It is essential to mention that Elon Musk announced his mission of acquiring Twitter in April 2022. In October 2022, he finalised the acquisition. We are also aware that some policies of Twitter have changed, such as content moderation, after the acquisition. During the time of acquisition, the event itself is toxic, the news outlets, and investor might have adjusted their engagement platform. This information hub, Twitter, is temporarily affected. Studies demonstrate that a majority ($60\%$) of Twitter users in the U.S. have taken a break from the platform since Elon's acquisition until May 2023, and that usage among the most active contributors has also reduced their activities after Musk’s takeover (see \cite{Hutchinson_TwitterMusk_2023}). Twitter was officially rebranded as X on July 23, 2023. This can also explain why the social media sentiment spillover network density is different from the news sentiment spillover network, and it decreases at the time of April 2022 and shows a downward trend again since the beginning of 2023, where the information transmission is even less connected.

 We zoom in to check the stability of each network under each rolling window. First, we ignore the weights on the edges, and the eigenvalues for each network are bigger than $1$, which indicates the networks will diverge over time. Second, we consider the original weights on the network, and compute the summation of the power matrix using equation \eqref{powermatrix}. We found that the network eventually becomes fully connected, indicating that the information is ultimately shared among all companies.  

\begin{figure}[H]
    \centering
    \begin{subfigure}{0.5\textwidth}
        \centering
        \includegraphics[width=\textwidth]{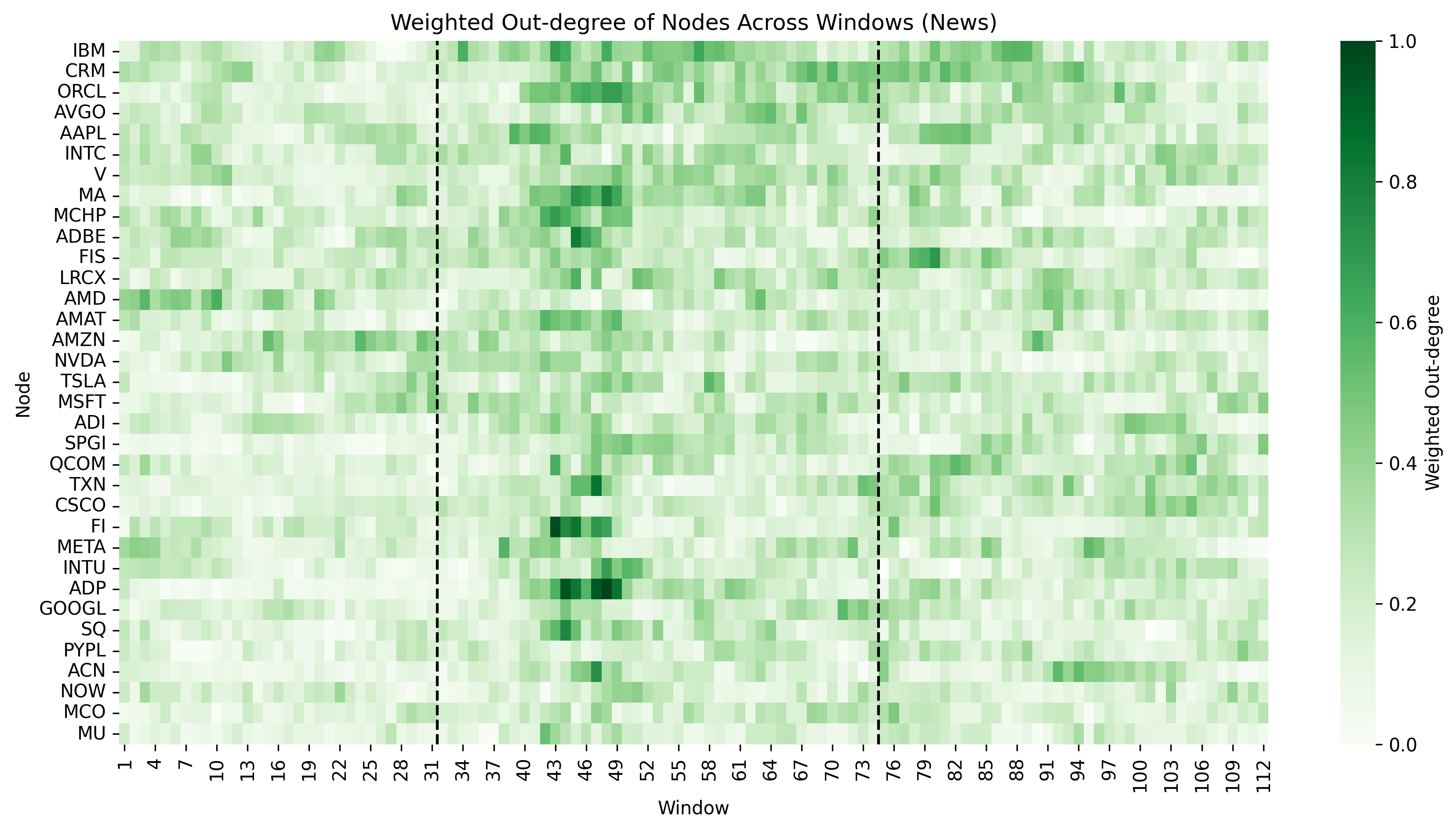}
        \caption{Out-degree heatmap (News)}
    \end{subfigure}
    \hfill
    \begin{subfigure}{0.49\textwidth}
        \centering
        \includegraphics[width=\textwidth]{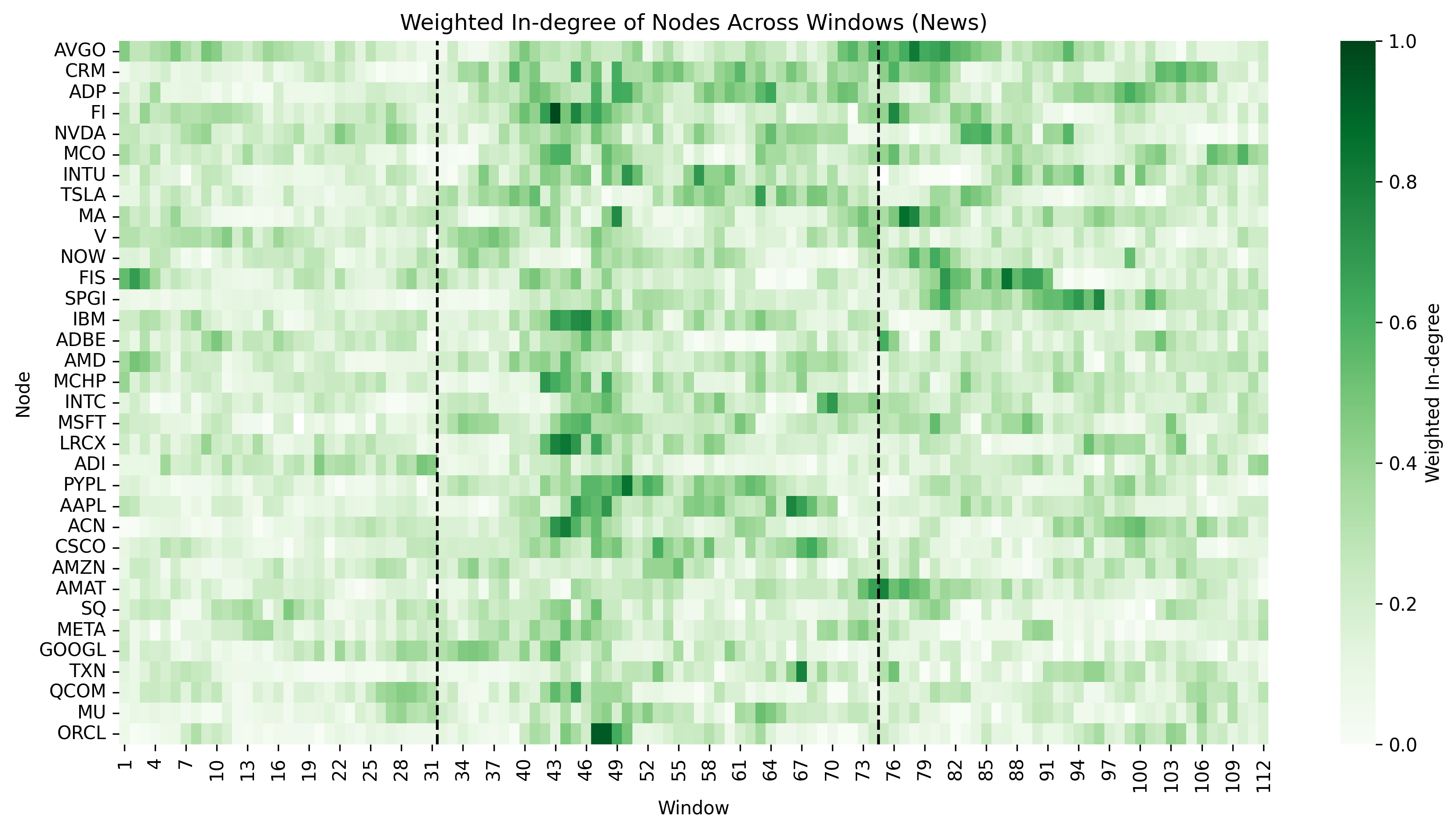}
        \caption{In-degree heatmap (News)}
    \end{subfigure}
    \caption{Out-degree and In-degree heatmaps of news sentiment network.}\caption*{ \small Notes: This figure shows the out-degree and in-degree intensity of companies across the news rolling windows. The black highlighted lines correspond to the spike period observed in the density plot in Figure \ref{te_sig_density}. Companies are ranked on the y-axis based on the frequency of their persistent influence over time. }
    \label{fig:news_heatmaps}
\end{figure}

\begin{figure}[H]
    \centering
    \begin{subfigure}{0.495\textwidth}
        \centering
        \includegraphics[width=\textwidth]{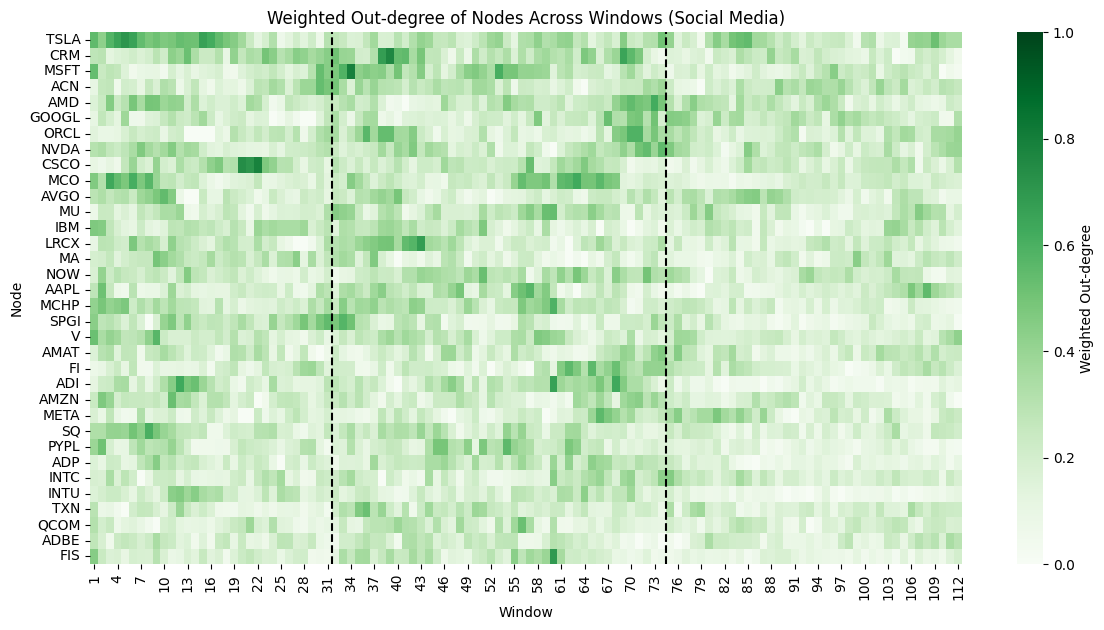}
        \caption{Out-degree heatmap (Social Media)}
    \end{subfigure}
    \begin{subfigure}{0.495\textwidth}
        \centering
        \includegraphics[width=\textwidth]{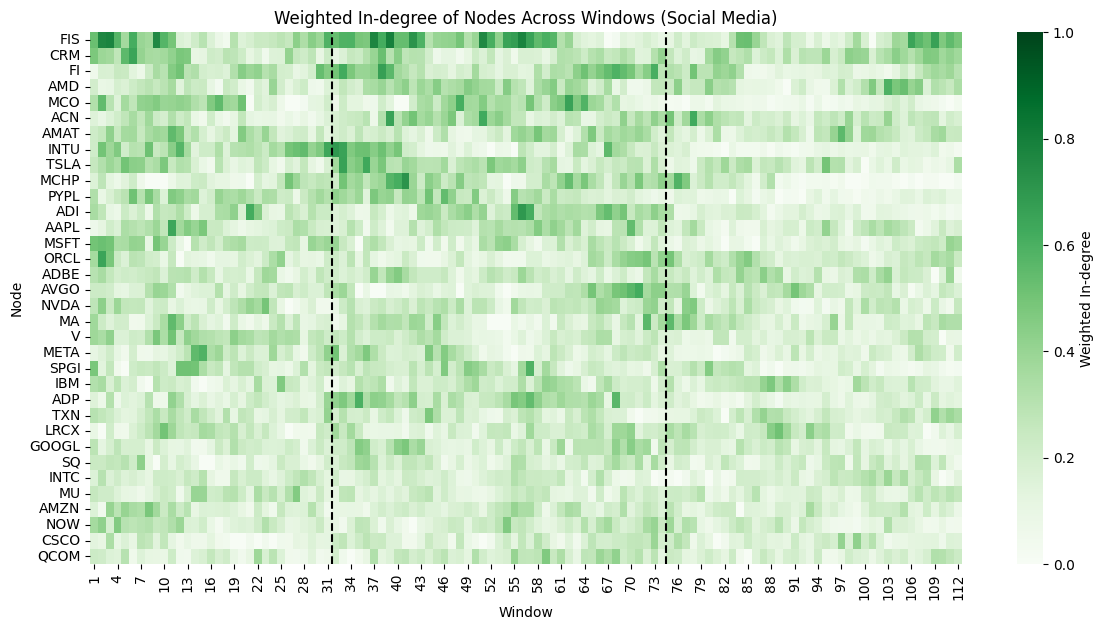}
        \caption{In-degree heatmap (Social Media)}
    \end{subfigure}
    \caption{Out-degree and in-degree heatmaps of the Social Media sentiment network.} \caption*{ \small Notes: Similarly, this figure shows the out-degree and in-degree intensity of companies across the social media rolling windows. The black highlighted lines correspond to the news highlighted period. Companies are ranked on the y-axis based on the frequency of their persistent high influence over time.}
    \label{fig:twitter_heatmaps}
\end{figure}

In order to check how the news and social media networks evolve over time, and which companies play an important role in spreading the information and absorbing the information. We provide the following heatmap plots: Figure \ref{fig:news_heatmaps} and \ref{fig:twitter_heatmaps}. In the plot, darker colours indicate a higher weighted in/out-degree of the nodes. Therefore, we can identify the most important companies in terms of both spreading and receiving information. We summarise the top 5 influential companies in Table \ref{chap4tab:pagedegree}, where we also include the PageRank results, which help us identify the most influential companies.

\begin{table}[H]
\centering
\caption{The top 5 influential companies}
\begin{tabularx}{\textwidth}{l|*{2}{X}|*{2}{X}|*{2}{X}}
\toprule
\textbf{Rank} & \multicolumn{2}{c|}{\textbf{PageRank}} & \multicolumn{2}{c|}{\textbf{Weighted Out-degree}} & \multicolumn{2}{c}{\textbf{Weighted In-degree}} \\
             & News & Social Media & News & Social Media & News & Social Media \\
\midrule
1 & AVGO  & FIS   & IBM  & TSLA  & AVGO & FIS   \\
2 & CRM  & FI    & CRM & CRM   & CRM & CRM   \\
3 & ADP  & CRM   & ORCL & MSFT  & ADP  & FI    \\
4 & FI & AMAT  & AVGO & ACN   & FI   & AMD   \\
5 & NVDA  & V     & AAPL  & AMD   & NVDA & MCO   \\
\bottomrule
\end{tabularx}\\
{\raggedright\small Note: This table provides the top 5 companies which show a persistent in-degree and out-degree influence of information flow based on the PageRank algorithm, weighted out-degree and weighted in-degree distribution.\par}
\label{chap4tab:pagedegree}
\end{table}
During the spike period in the news network, we observe in the heatmap plot (Figure \ref{fig:news_heatmaps} and \ref{fig:twitter_heatmaps}) that the intensity of the information flow increased for most of the companies, whereas the pattern is more random in Social Media sentiment networks, but there is a clear trend that the intensity faded following Twitter's rebranding. 

Notably, Salesforce (CRM) appears the most frequently, which shows a strong and consistent intensity in transferring and absorbing information. It is the world-leading customer relationship management platform, widely used by companies for managing sales, marketing, etc. Similarly, Broadcom (AVGO), as one of the biggest companies globally and a technology leader, provides both hardware and software. By the time of writing \citep{Investopedia_Mag7_2025}, its significantly growing performance makes it a competitor to Tesla (TSLA) to gain a seat in the Magnificent Seven companies, but our analysis could signal this by anchoring its information spillover position using historical data. Specifically, in the news sentiment spillover network, International Business Machines Corporation (IBM), CRM, Oracle (ORCL) and Apple (AAPL), demonstrate a persistent leading role in spreading the information. In contrast, Nvidia's (NVDA) presence is closely tied to the AI boom, making it highly sensitive to market news. It is not surprising that TSLA acts as the king of social media influence. Social Media often reflects market hype and serves as a collective source of speculative content. Moody's (MCO), as a credit rating agency, operates based on the climate of the companies and markets. This makes them a natural information receiver in social media. We can also notice that Fidelity (FIS) and Fiserv (FI) are major information receivers. These companies provide payment systems to tech companies, making their sentiment closely linked to market speculation surrounding the tech sector, similar to the case of Moody's.
 
Overall, we believe that the news and social media sentiment spillover network is different over time, and sometimes independently of the major market events. The information hub companies are also quite different from each other. In general, we think news tends to diffuse faster among companies, which is consistent with the findings of \cite{hou_industry_2007}. In addition,
online users tend to focus on the individual company, which can be chaotic, and there are also many disagreements on social media \citep{reboredo_impact_2018, cookson_investor_2023}. Contents from online platforms (Stocktwits, Twitter) are more emotion-driven collective influence, whereas news platform sentiment is often driven by macroeconomic factors and the company itself, suggesting a more selective form of influence. 

\subsection{Three regimes network analysis}

We further identify three different sentiment spillover regimes from the news network density plot, which are more likely to reflect shifts in investor behaviour towards the US technology industry. The three regime periods are before 2021-08-02, 2021-08-02 to 2022-10-03 and after 2022-10-03. To make the comparison consistent, we apply the same regime periods to Social Media. 
\subsubsection{News}

\begin{figure}[H]
        \centering
        \begin{subfigure}[b]{0.32\textwidth}
            \includegraphics[width=\linewidth]{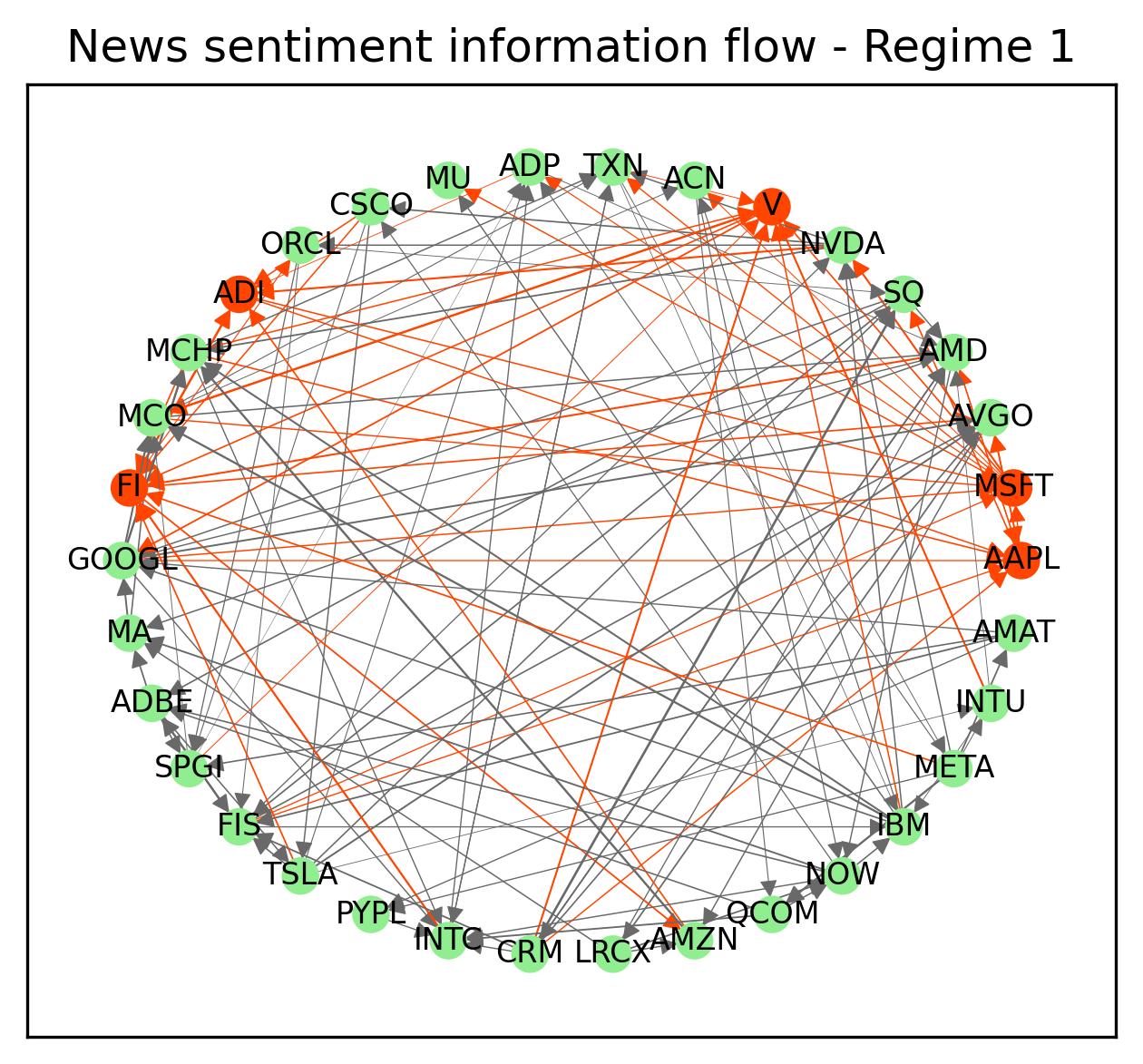}
            \caption{Before 2021-08}
        \end{subfigure}
        \hfill
        \begin{subfigure}[b]{0.32\textwidth}
            \includegraphics[width=\linewidth]{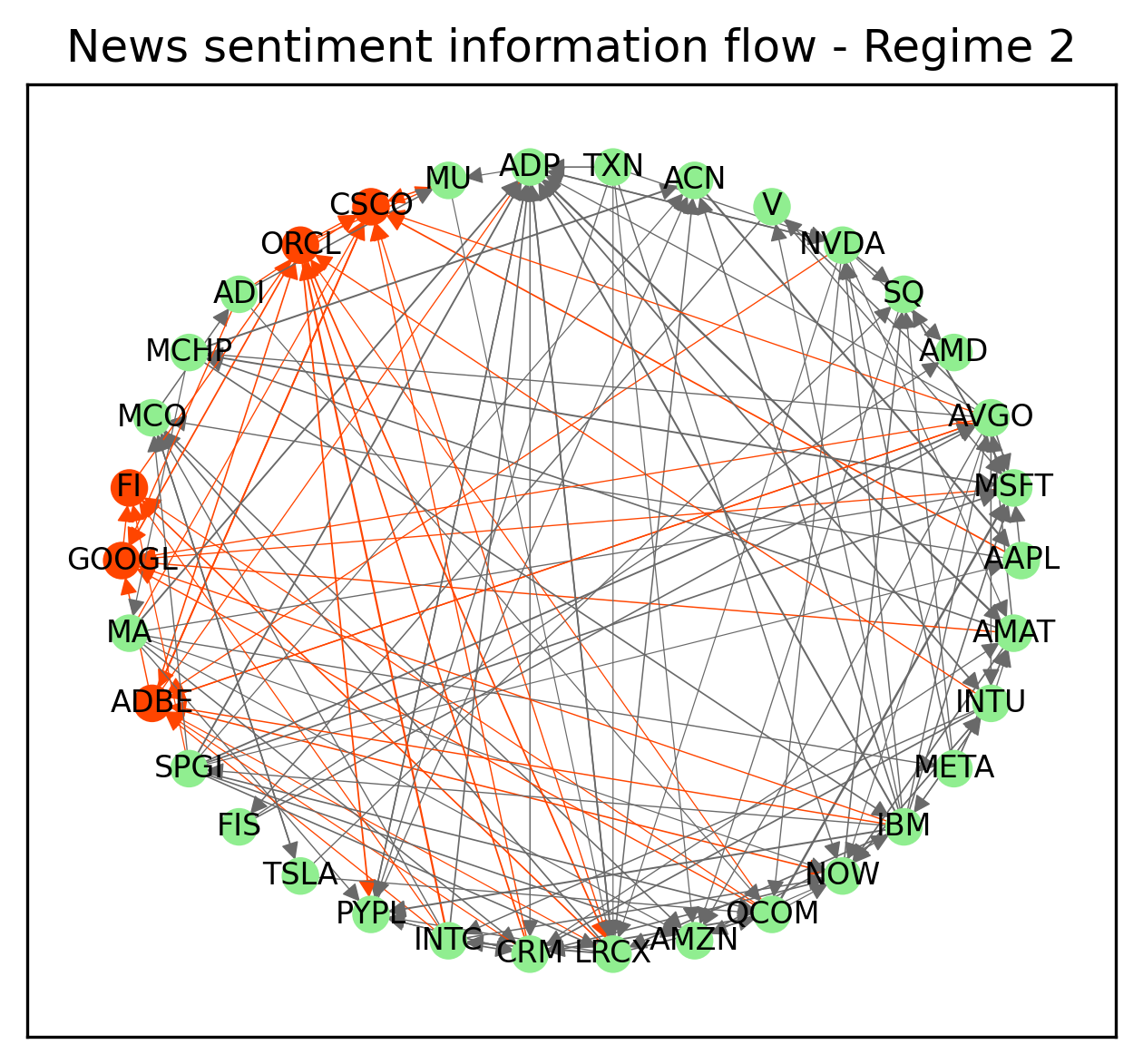}
            \caption{2021-08 to 2022-10}
        \end{subfigure}
        \hfill
        \begin{subfigure}[b]{0.32\textwidth}
            \includegraphics[width=\linewidth]{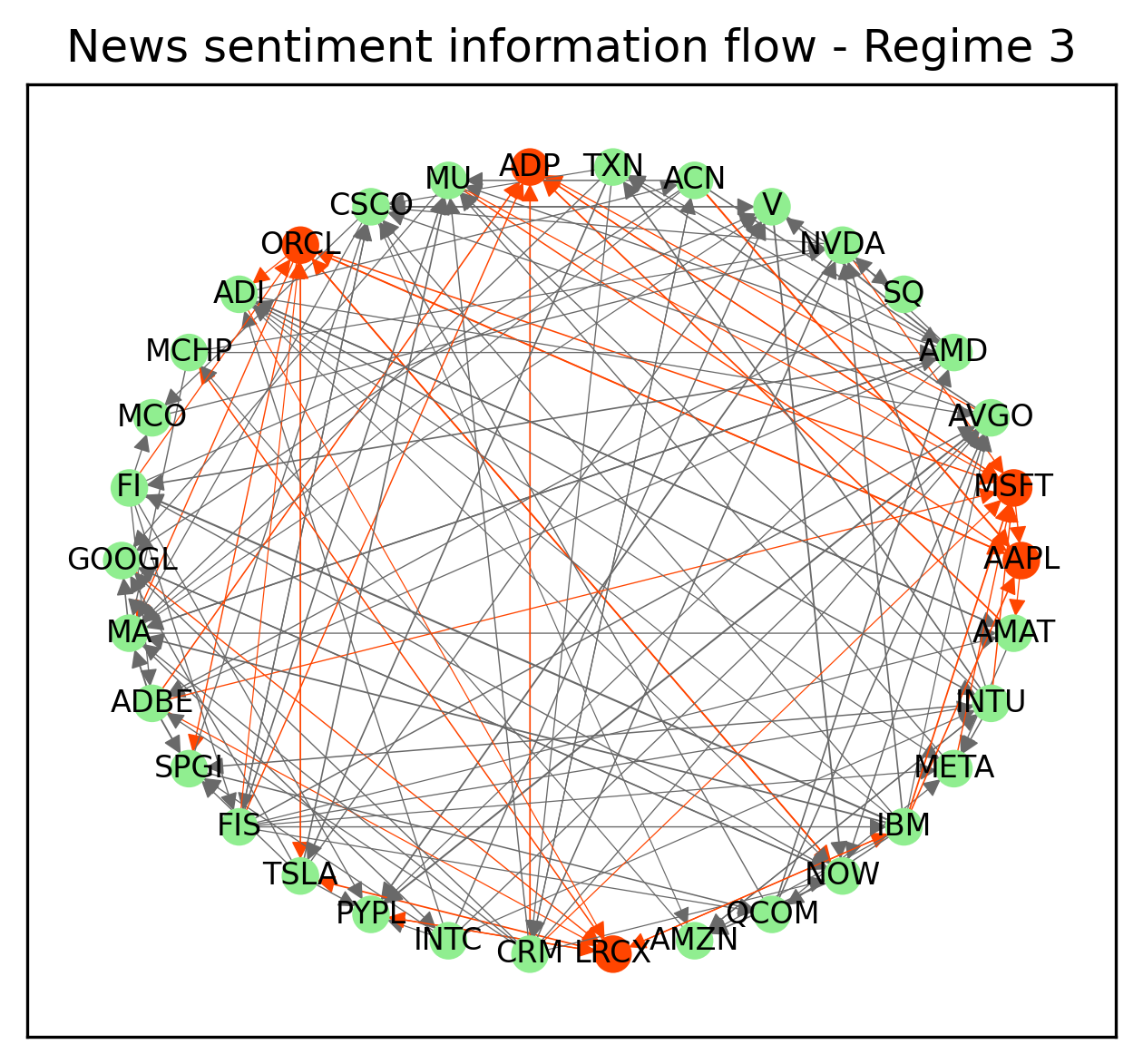}
            \caption{After 2022-10}
        \end{subfigure}
        \caption{News sentiment network visualisations under three regimes. }
        \caption*{\small Notes: The red-highlighted nodes are selected as the top 5 influencers based on the PageRank algorithm. }
        \label{news_netviz}
    \end{figure}

\begin{figure}[H]
        \centering
        \begin{subfigure}[b]{0.33\textwidth}
            \includegraphics[width=\linewidth]{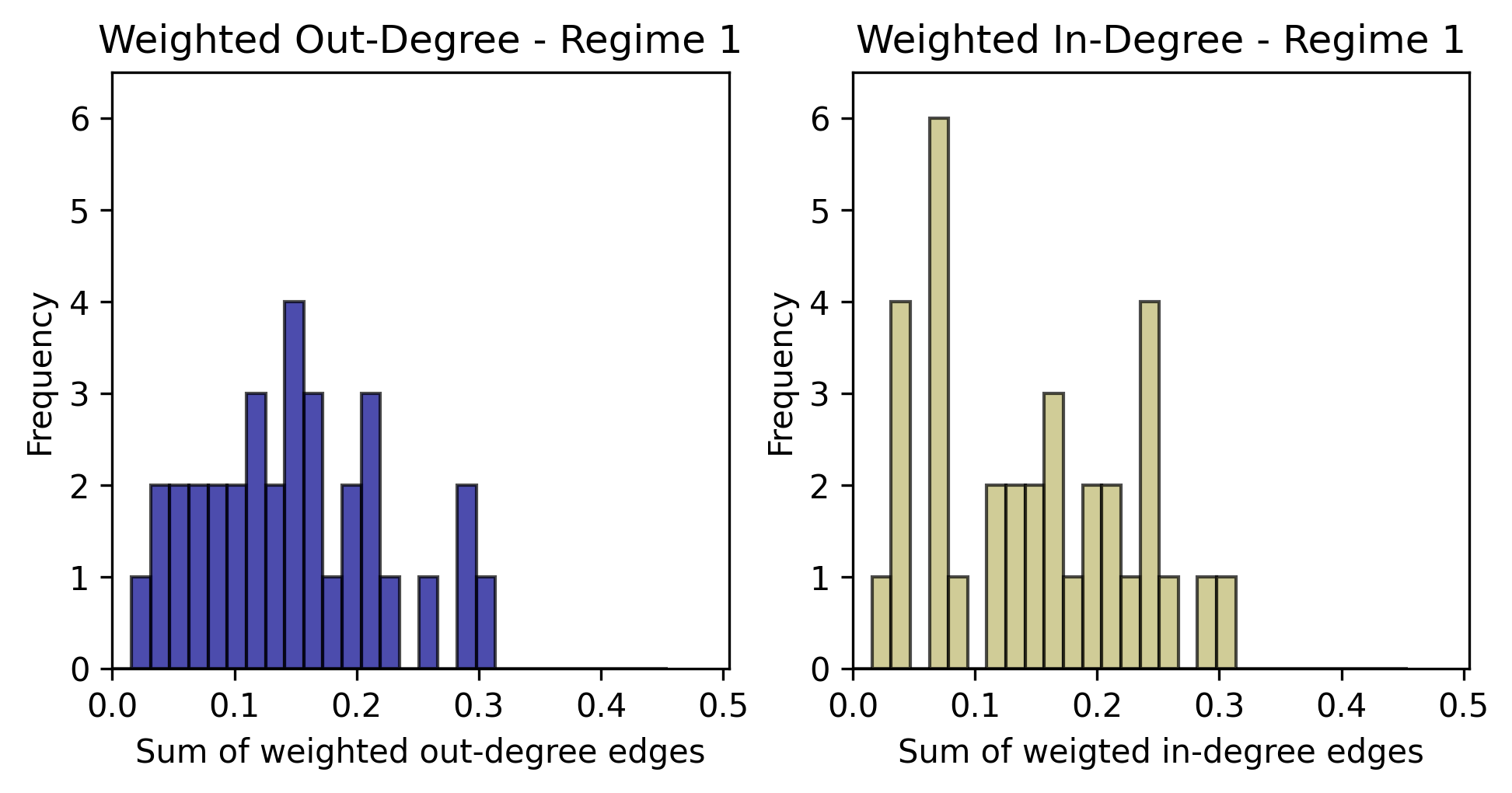}
            \caption{Before 2021-08}
        \end{subfigure}
        \hfill
        \begin{subfigure}[b]{0.33\textwidth}
            \includegraphics[width=\linewidth]{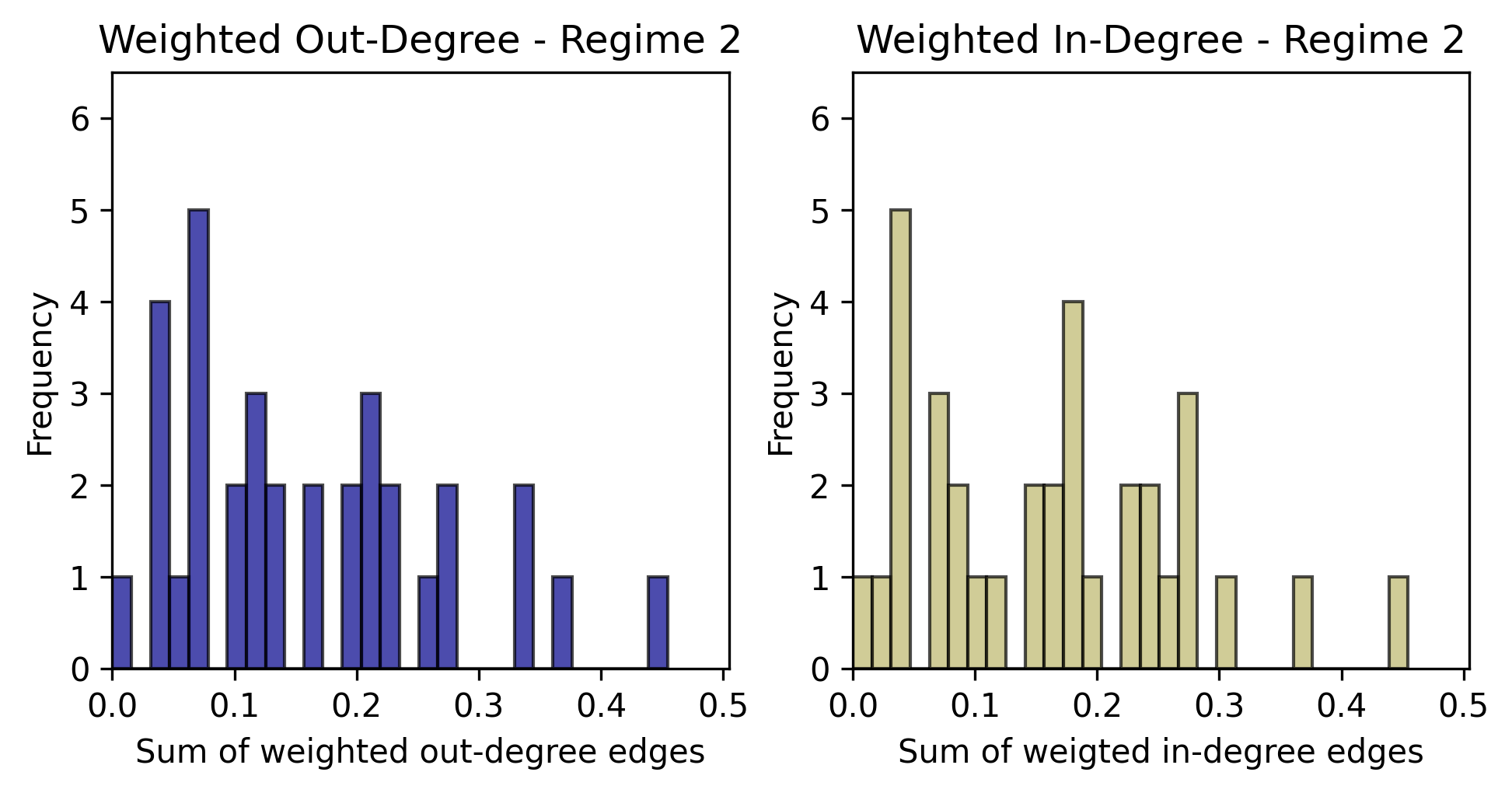}
            \caption{2021-08 to 2022-10}
        \end{subfigure}
        \hfill
        \begin{subfigure}[b]{0.32\textwidth}
            \includegraphics[width=\linewidth]{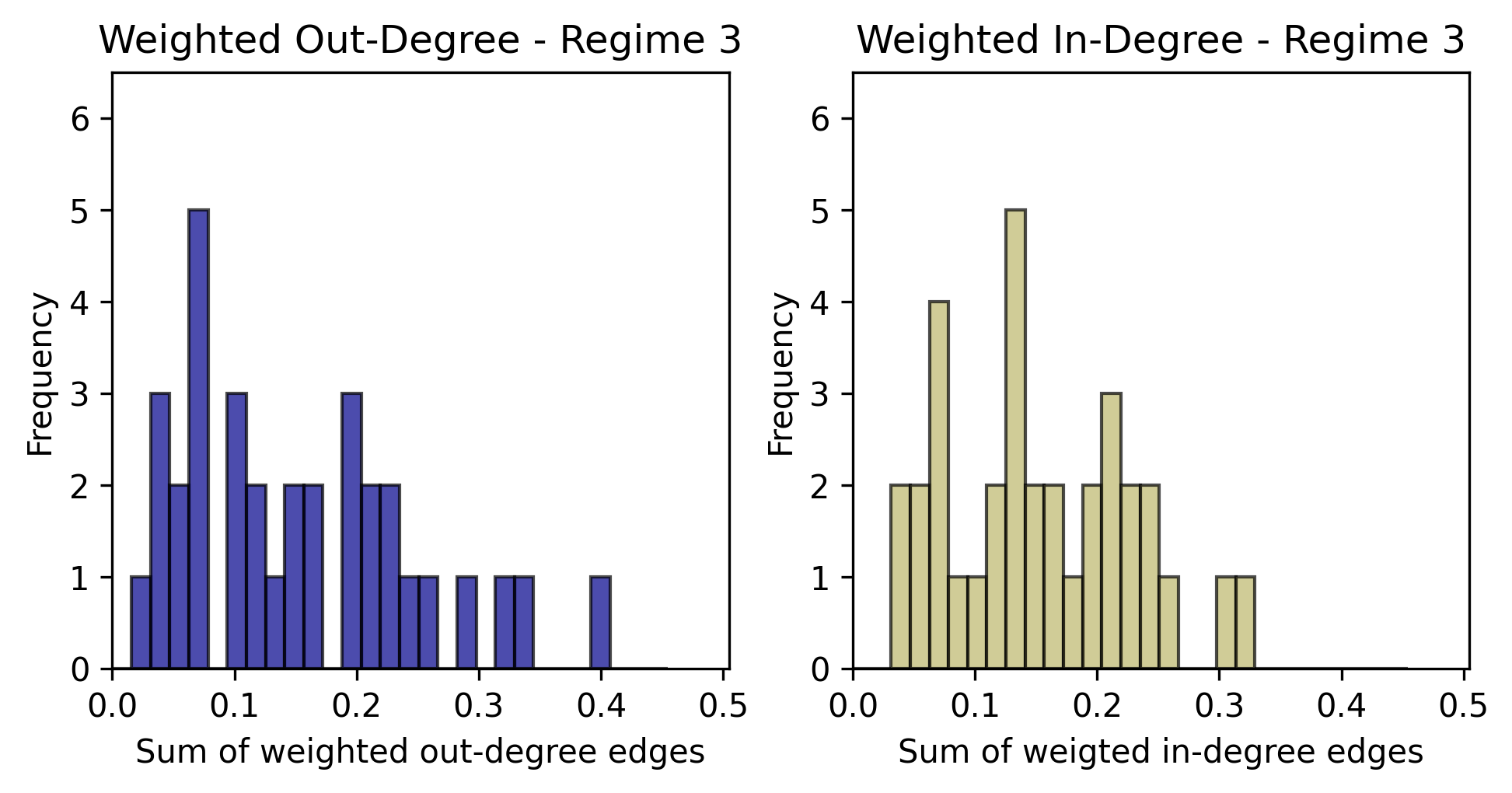}
            \caption{After 2022-10}
        \end{subfigure}
        \caption{Weighted in-degree and out-degree distribution (News).}\caption*{\small Notes: This figure shows the weighted in-degree and out-degree distributions for the news sentiment networks under different regimes.}
        \label{degreedistri}
    \end{figure}

In the news sentiment information transfer entropy network visualisations (Figure \ref{news_netviz}) for three different periods, where the nodes represent different companies, and the weighted edges represent the information flow between companies. It is clear to see that the second period, which corresponds to the evident increase in the network density plot in Figure \ref{te_sig_density}, has the most dense information transfer. This trend continues, and the third period has more edges than the first period. This can also be seen in the degree distribution plot in Figure \ref{degreedistri}. The degree distributions show a right-skewed pattern. Most of the edge values are centred in a lower range, with a long tail. During the second period, there is a much wider spread in both weighted in-degree and out-degree, suggesting a stronger information transmission. Some nodes with high sum weights suggest their importance in spreading the information, with over 0.4 weighted edges. 
Regime 1 has a relatively narrow range, with most nodes clustered around 0.05 to 0.3, with one or two companies acting as influencers. While regime 3 seems to have a wider range than regime 1, but with more nodes have a higher sum of weights, which could be the re-stabilisation after regime 2. 

In the table below (Table \ref{tab:centrality_by_regime}), we rank the company in-degree and out-degree centrality and list the top 5 companies under three different periods. IBM appears in three regimes, and it has a high weighted out-degree centrality. IBM is a leading provider of global software, cloud services, AI and consulting, which indicates the development of IBM may affect other companies in the tech sector. The result is in line with \citep{nyakurukwa_investor_2025}, who explored the news sentiment connectedness among Dow Jones Industrial Average (DJIA) index firms, which consist of blue chip companies, and concluded that IBM acts consistently as a net transmitter of news sentiment. We notice that financial services companies tend to behave as information receivers. 
We further take a look at which companies contribute more to shifting the overall network density. It turns out that semiconductor companies, such as Lam Research (LRCX), Advanced Micro Devices (AMD), and AVGO, contribute the most in shifting the information flow. One of the reasons is that since COVID, the demand for semiconductors has expanded due to the fundamental technological advancement in semiconductors \citep{CPRAM_TechCycle_2024}. In 2021, semiconductor market revenue worldwide increased by around $22\%$, which is the highest increase within the study period from 2019 to 2024. 

\begin{table}[H]
\caption{In-degree and Out-degree Centrality by Regime}
\label{tab:centrality_by_regime}
\begin{tabularx}{\textwidth}{lXXl}
\toprule
\textbf{Regime 1} & & & \\
\midrule
\textbf{Company} & \textbf{In-degree} & \textbf{Company} & \textbf{Out-degree} \\
FI (FS)   & 0.302 & CRM (SW)   & 0.301 \\
V (FS)    & 0.298 & \textcolor{red}{GOOGL (CS)} & 0.291 \\
AVGO (SC) & 0.256 & IBM (IT)   & 0.284 \\
INTC (SC) & 0.245 & \textcolor{red}{META (CS)}  & 0.261 \\
AMD (SC)  & 0.245 & FIS (FS)   & 0.224 \\
\midrule

\textbf{Regime 2} & & & \\
\midrule
\textbf{Company} & \textbf{In-degree} & \textbf{Company} & \textbf{Out-degree} \\
ADP (PS)   & 0.450 & IBM (IT)   & 0.455 \\
\textcolor{red}{MSFT (SW)}  & 0.371 & SPGI (CM)  & 0.373 \\
CSCO (CE)  & 0.313 & LRCX (SC)  & 0.344 \\
\textcolor{red}{GOOGL (CS)} & 0.280 & AVGO (SC)  & 0.341 \\
NOW (SW)   & 0.277 & MA (FS)    & 0.274 \\
\midrule

\textbf{Regime 3} & & & \\
\midrule
\textbf{Company} & \textbf{In-degree} & \textbf{Company} & \textbf{Out-degree} \\
MA (FS)   & 0.322 & IBM (IT)   & 0.403 \\
\textcolor{red}{MSFT (SW)} & 0.313 & CRM (SW)   & 0.331 \\
PYPL (FS) & 0.255 & FIS (FS)   & 0.326 \\
MU (SC)   & 0.249 & AMD (SC)   & 0.285 \\
NOW (SW)  & 0.247 & ORCL (SW)  & 0.260 \\
\bottomrule
\end{tabularx}
\small{Note: This table displays the top five companies with the highest centrality scores in each regime, with GICS sector abbreviations included. The Magnificent Seven companies are highlighted in red to clearly indicate their positions in each regime.}
\end{table}

\begin{figure}[H]
        \centering
        \begin{subfigure}[b]{0.32\textwidth}
            \includegraphics[width=\linewidth]{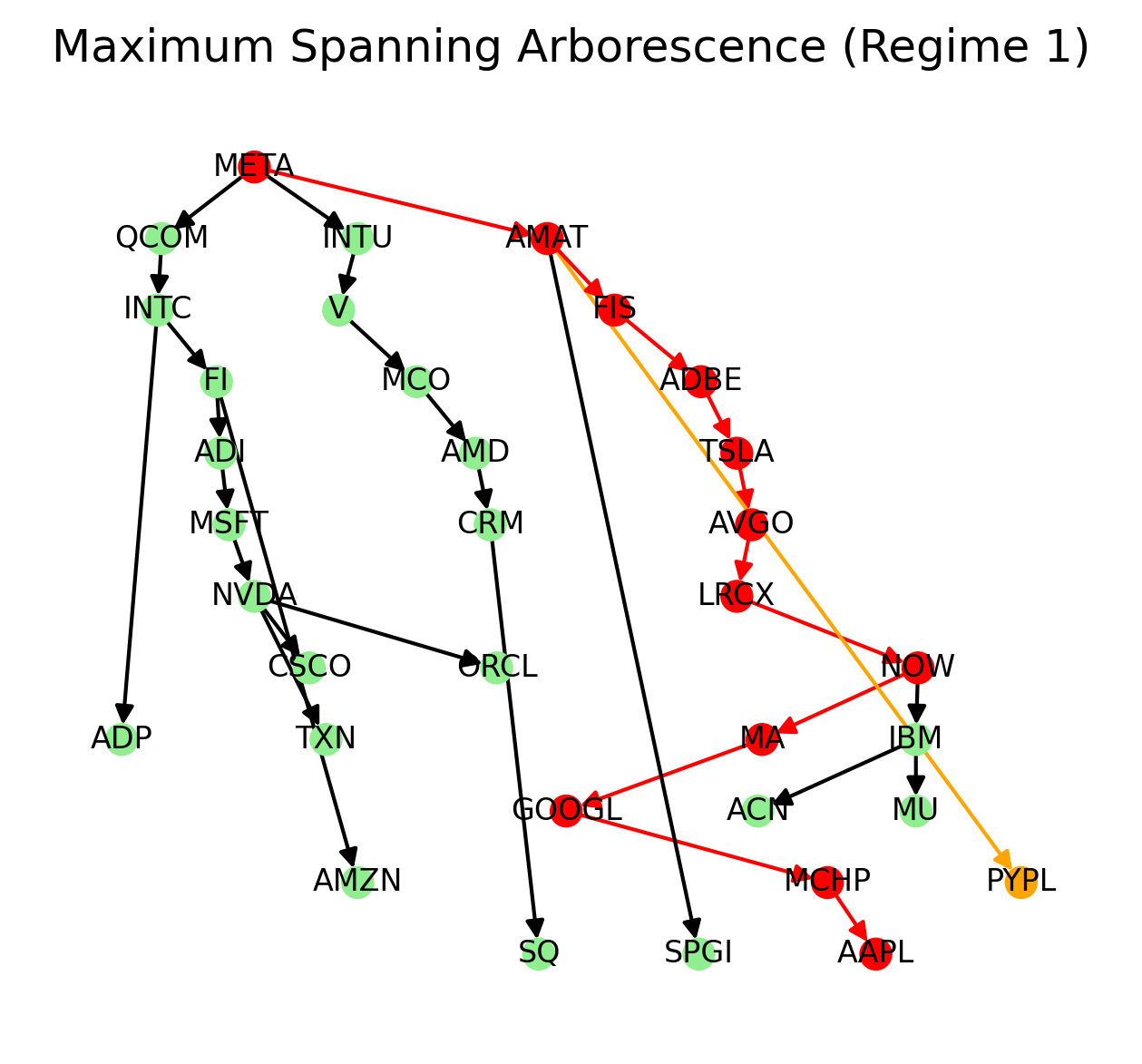}
            \caption{Before 2021-08}
        \end{subfigure}
        \hfill
        \begin{subfigure}[b]{0.32\textwidth}
            \includegraphics[width=\linewidth]{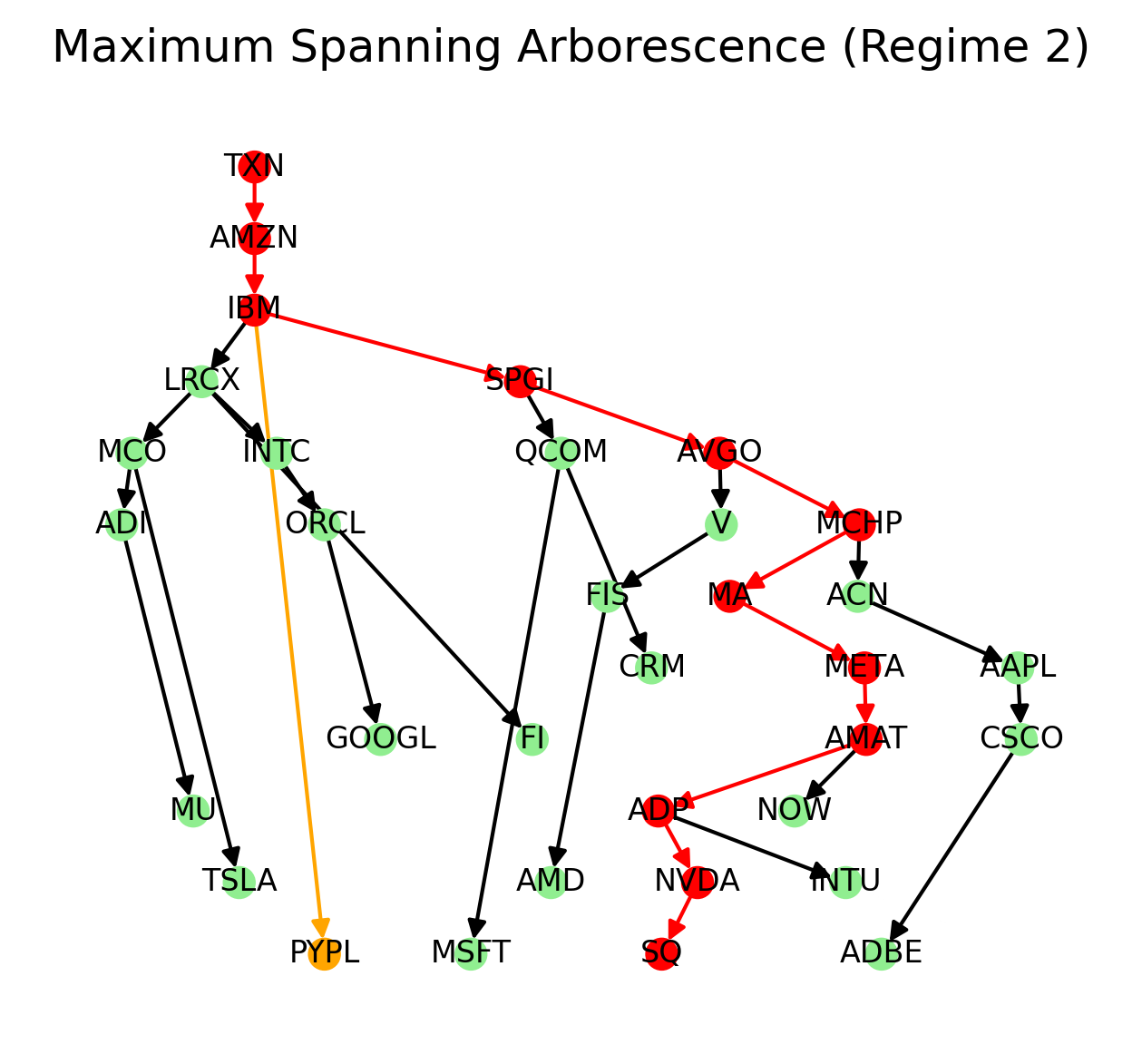}
            \caption{2021-08 to 2022-10}
        \end{subfigure}
        \hfill
        \begin{subfigure}[b]{0.32\textwidth}
            \includegraphics[width=\linewidth]{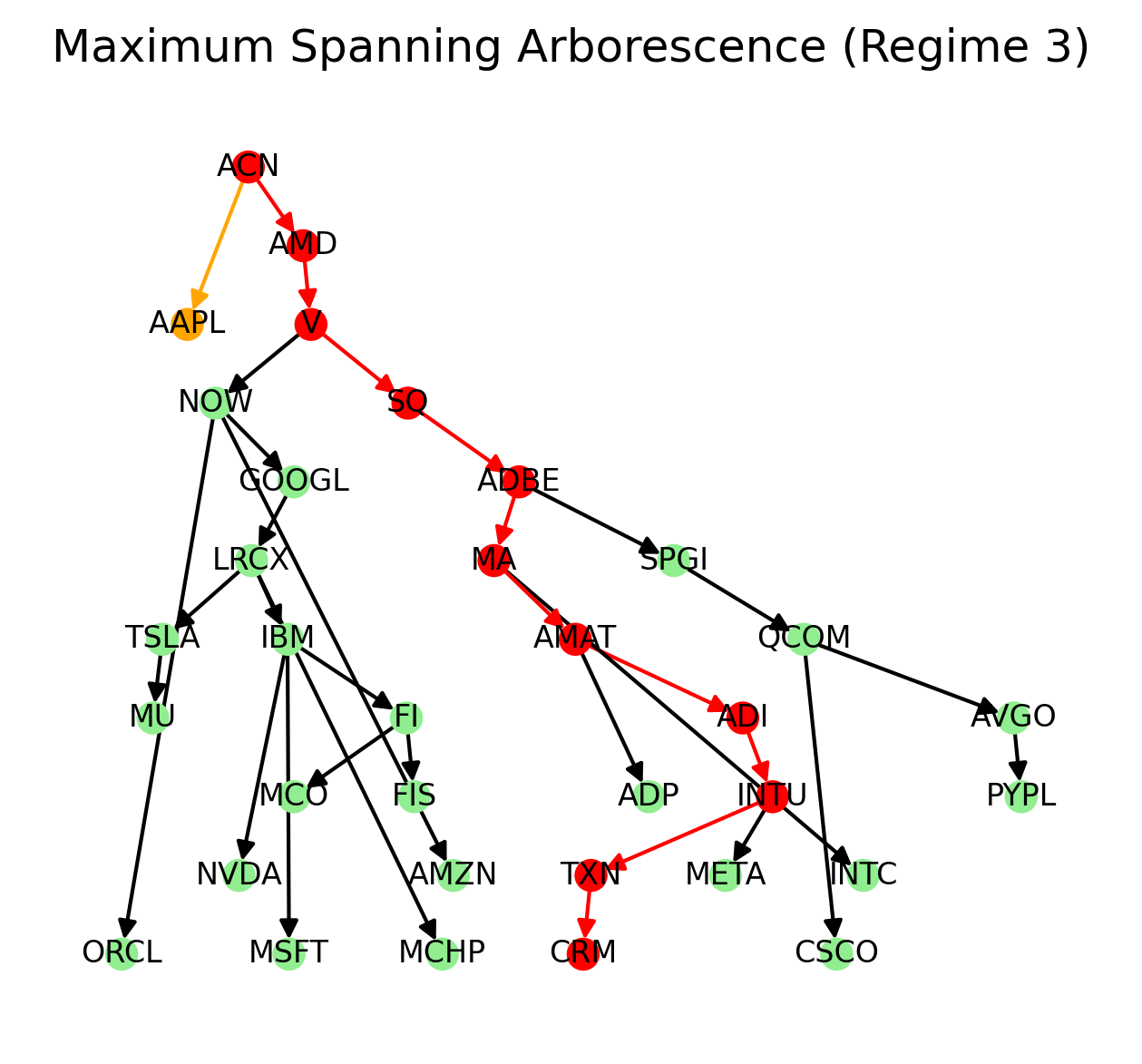}
            \caption{After 2022-10}
        \end{subfigure}
        \caption{Maximum Spanning Arborescence (News)}\caption*{ \small Notes: This figure displays the maximum spanning arborescence structure for three regimes in the news information spillover network. The red-highlighted path is the strongest weighted path, and the lowest weighted path is highlighted in orange. }
        \label{msa news}
    \end{figure}

\begin{figure}[H]
    \centering
    \begin{subfigure}{0.49\linewidth}
        \centering
        \includegraphics[width=\linewidth]{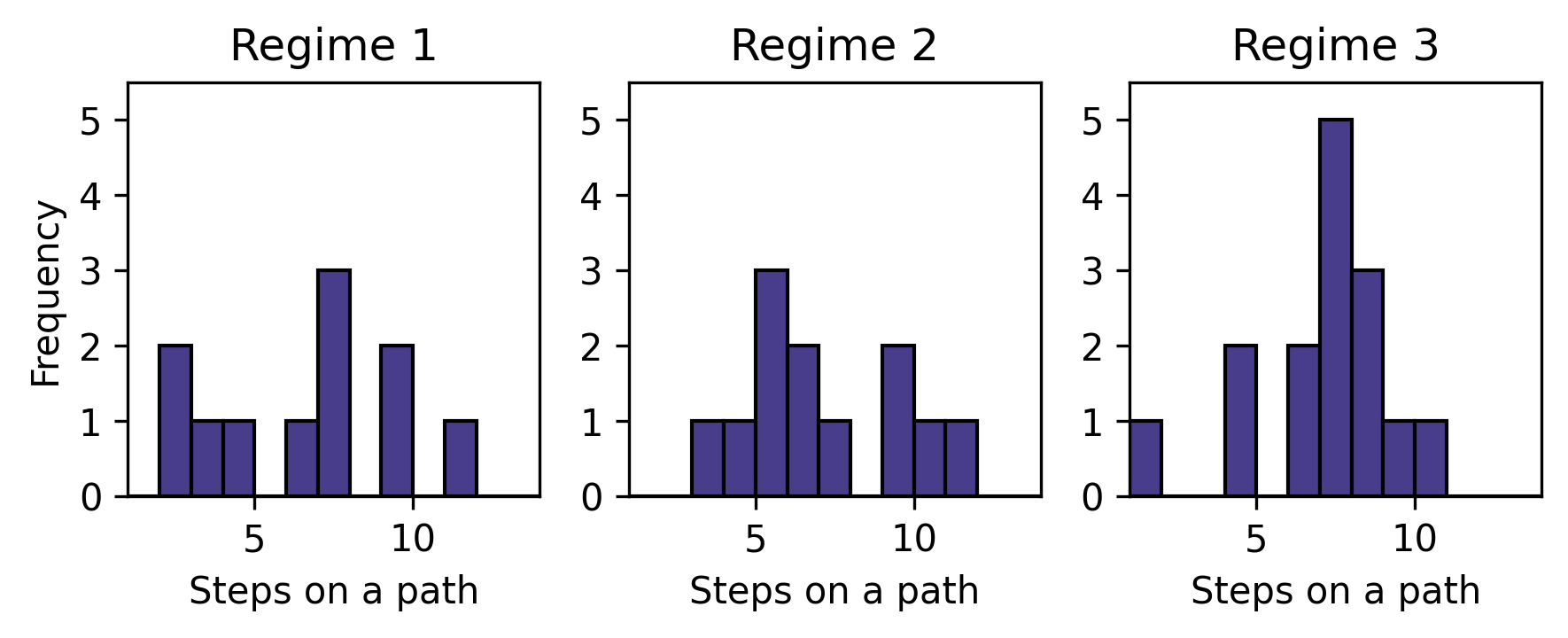}
        \caption{Number of steps}
        \label{fig:steps_distribution}
    \end{subfigure}
    \hfill
    \begin{subfigure}{0.49\linewidth}
        \centering
        \includegraphics[width=\linewidth]{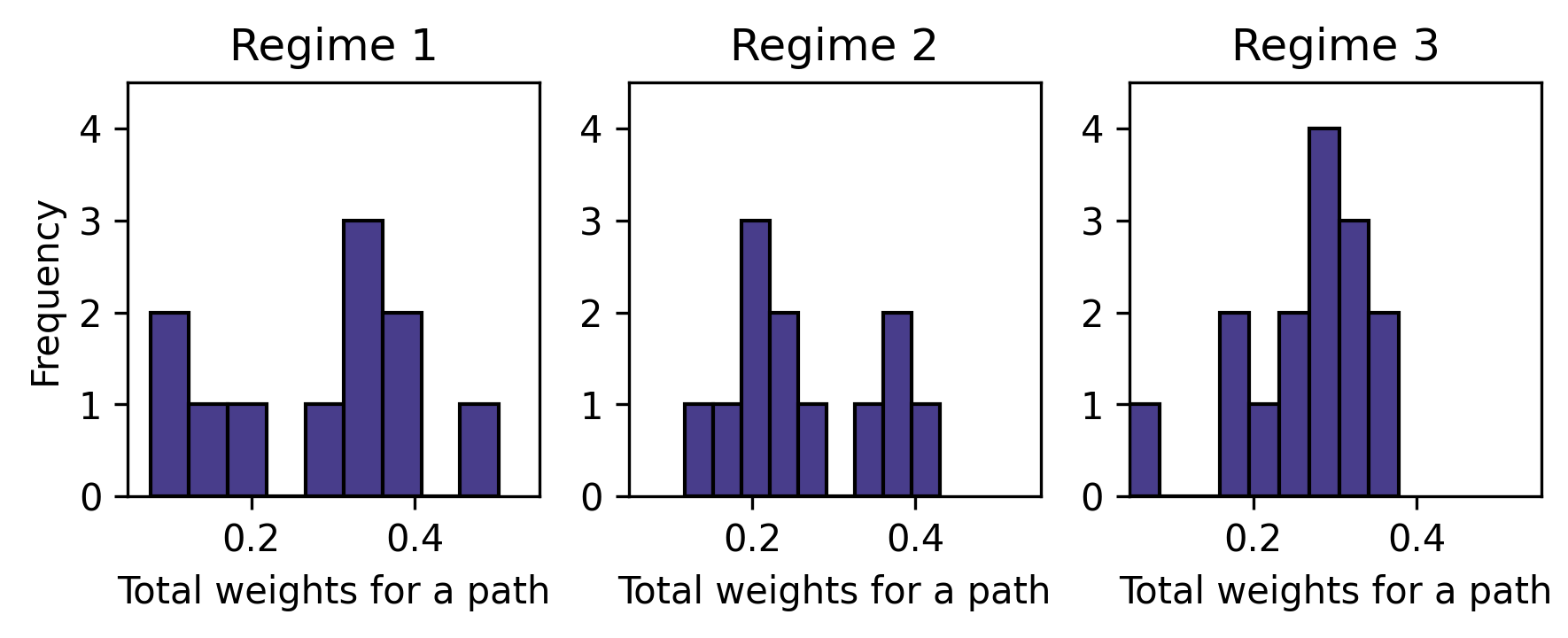}
        \caption{Total weights}
        \label{fig:length_distribution}
    \end{subfigure}
\caption{Distribution plot of the number of steps and total weights on a path.}
\end{figure}

Furthermore, we estimate the Maximum Spanning Arborescence (MSA) for each regime to extract the maximum possible weighted edges in the network, which is the maximum possible dominant information diffusion paths. The results are shown in Figure \ref{msa news}. The total number of paths for these three regimes is 11, 12, and 15, respectively. The highest weighted path is highlighted in red, indicating the strongest information flow path. Relatively, the lowest weighted path is marked in orange. The distribution of the number of steps and the sum of the weights on a path is plotted in Figure \ref{fig:steps_distribution} and Figure \ref{fig:length_distribution}. Notably, the majority of paths have an average of 6 steps in regime 2, which is lower than in regimes 3 and 1, where we frequently observe more than 7 steps for a path. In regime 2, there is a clear broadcast pattern of the information flow path, indicating the information diffuses faster and more or less evenly, with a higher chance of information spillover. Whereas in regime 3 and regime 1, there is a long chain with many steps supporting the information to spread, showing a clear sequential ripple and prolonged effect in spreading the information. The total weight distribution for a path across three regimes also suggests that in regime 2, it is slightly more concentrated. Companies such as MA, META, AVGO, and ADBE appear frequently on the highest weighted information flow path, implying these companies act as critical bridges in receiving and spreading the information. 


\subsubsection{Social Media }
\begin{figure}[H]
        \centering
        \begin{subfigure}[b]{0.32\textwidth}
            \includegraphics[width=\linewidth]{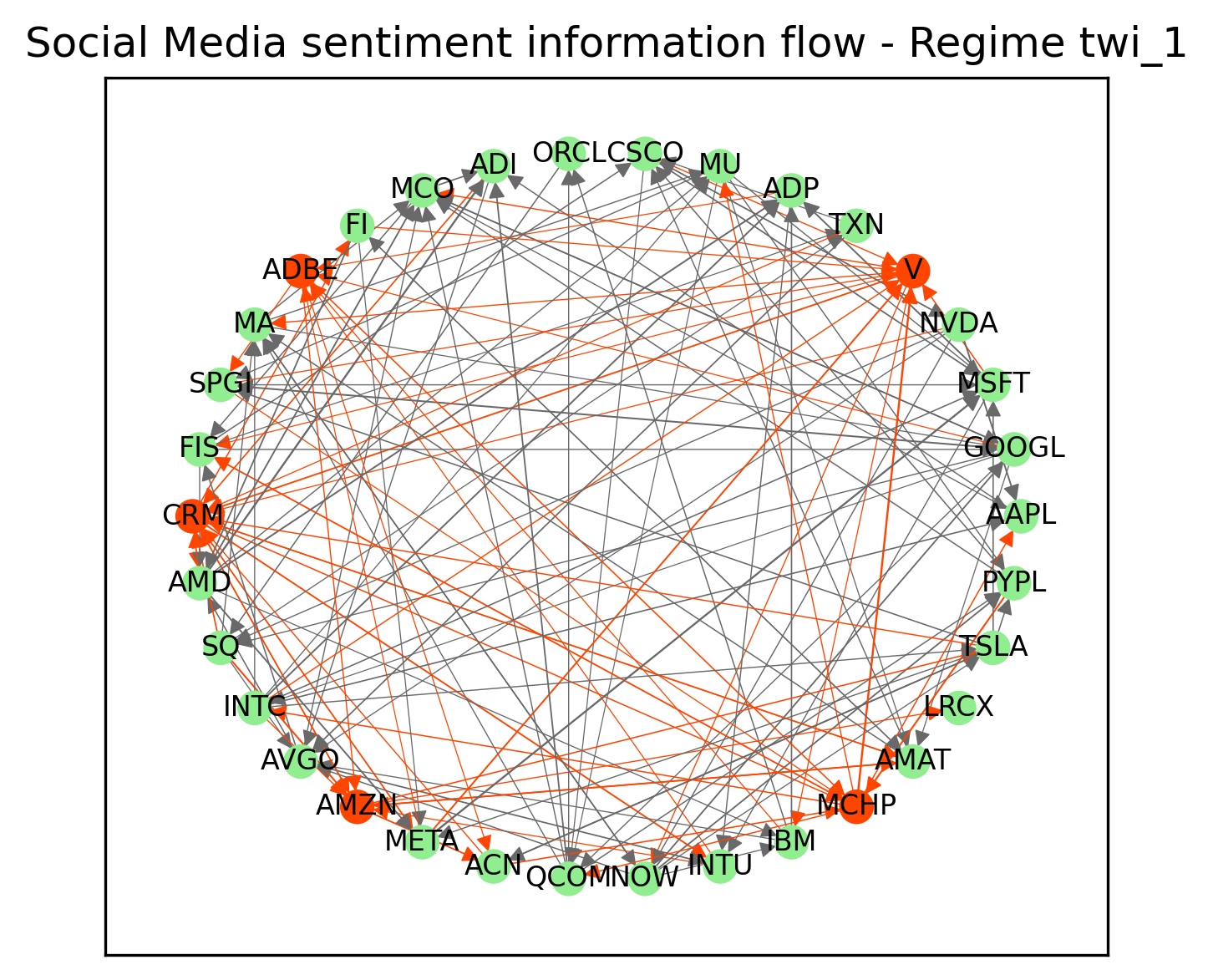}
            \caption{Before 2021-08}
        \end{subfigure}
        \hfill
        \begin{subfigure}[b]{0.32\textwidth}
            \includegraphics[width=\linewidth]{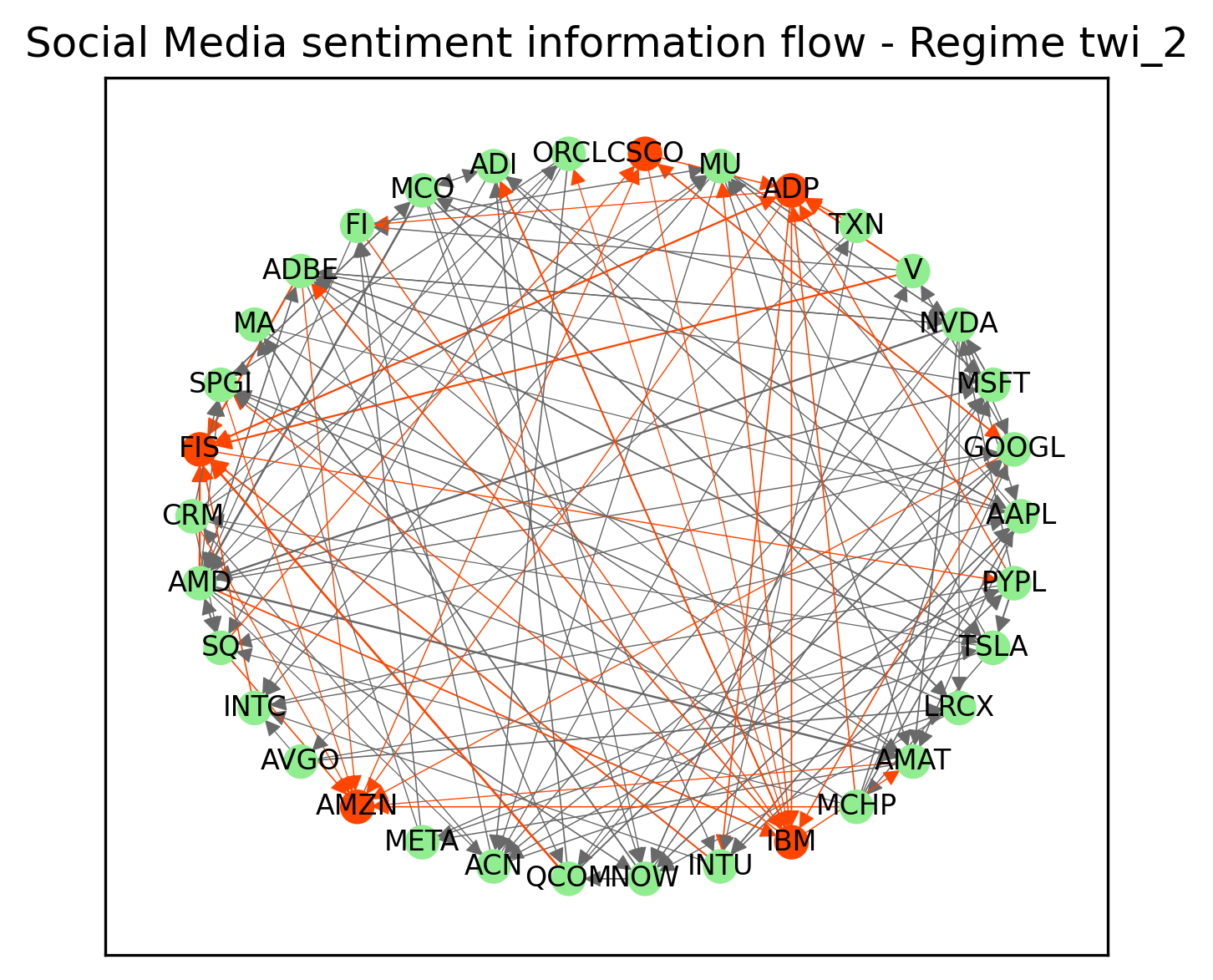}
            \caption{2021-08 to 2022-10}
        \end{subfigure}
        \hfill
        \begin{subfigure}[b]{0.32\textwidth}
            \includegraphics[width=\linewidth]{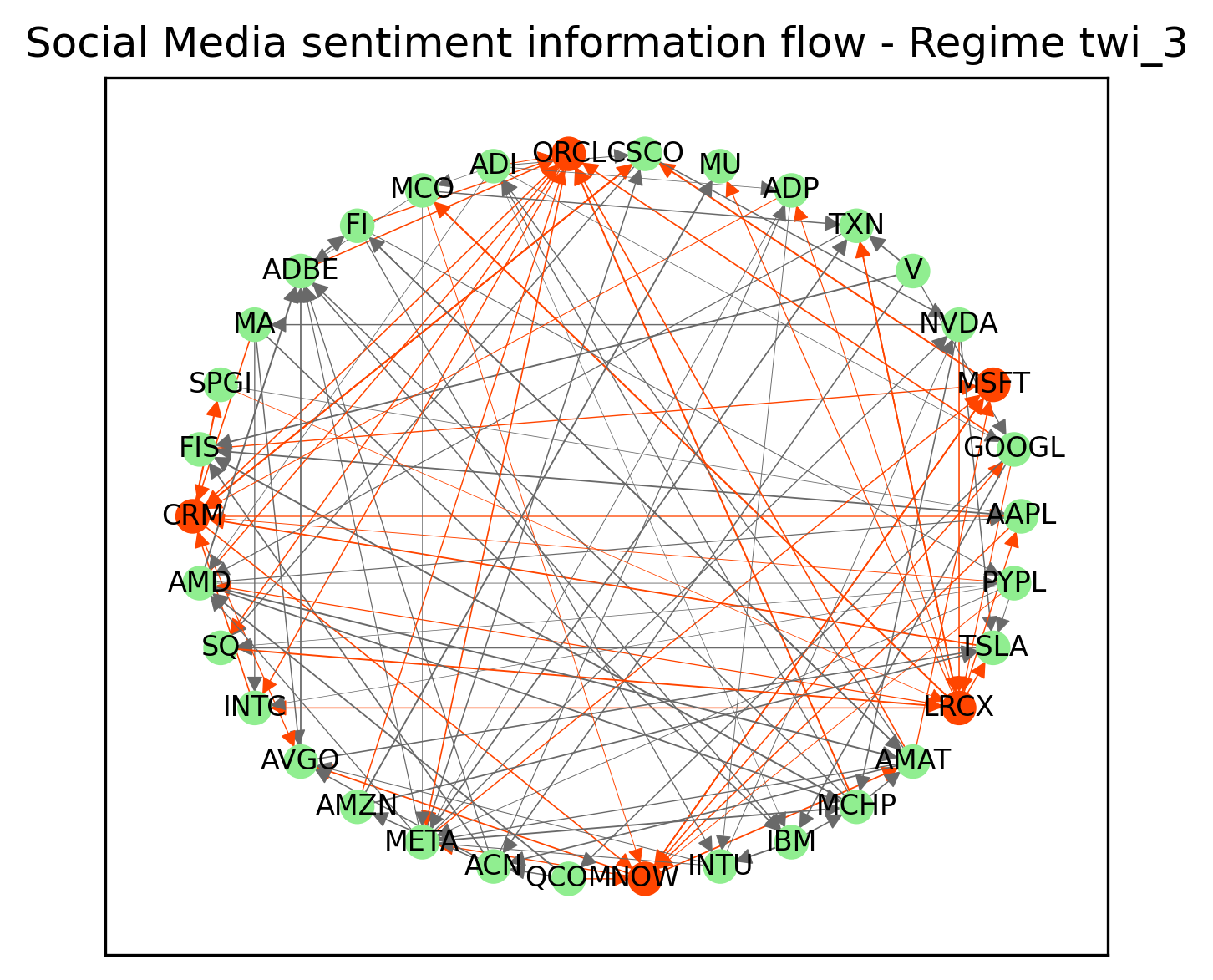}
            \caption{After 2022-10}
        \end{subfigure}
        \caption{\small Social media sentiment network visualisations under three regimes.}\caption*{\small Notes: The red-highlighted nodes are selected as the top 5 influencers based on the PageRank algorithm. }
        \label{regimes_twi}
    \end{figure}

\begin{figure}[H]
        \centering
        \begin{subfigure}[b]{0.32\textwidth}
            \includegraphics[width=\linewidth]{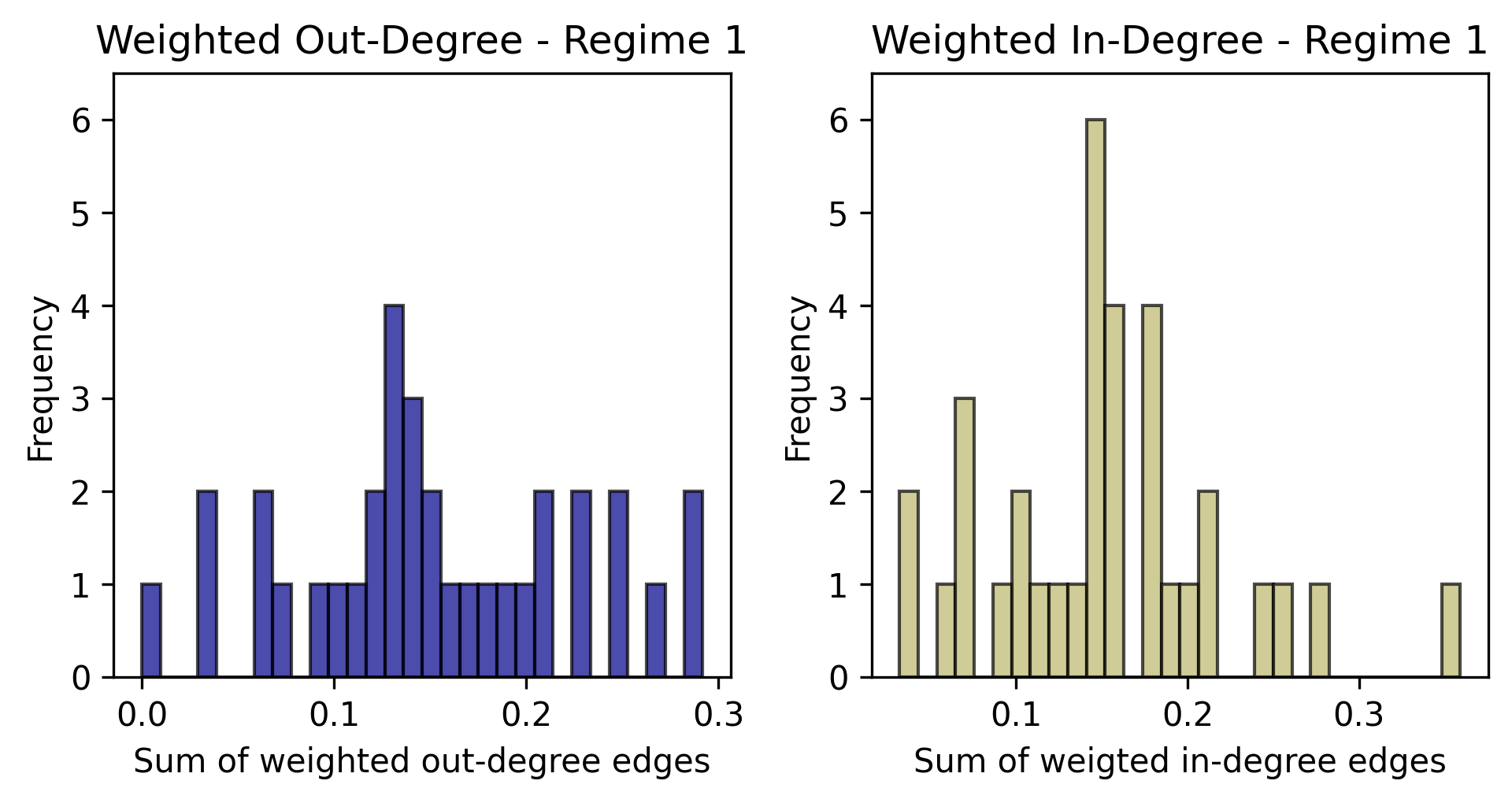}
            \caption{Before 2021-08}
        \end{subfigure}
        \hfill
        \begin{subfigure}[b]{0.32\textwidth}
            \includegraphics[width=\linewidth]{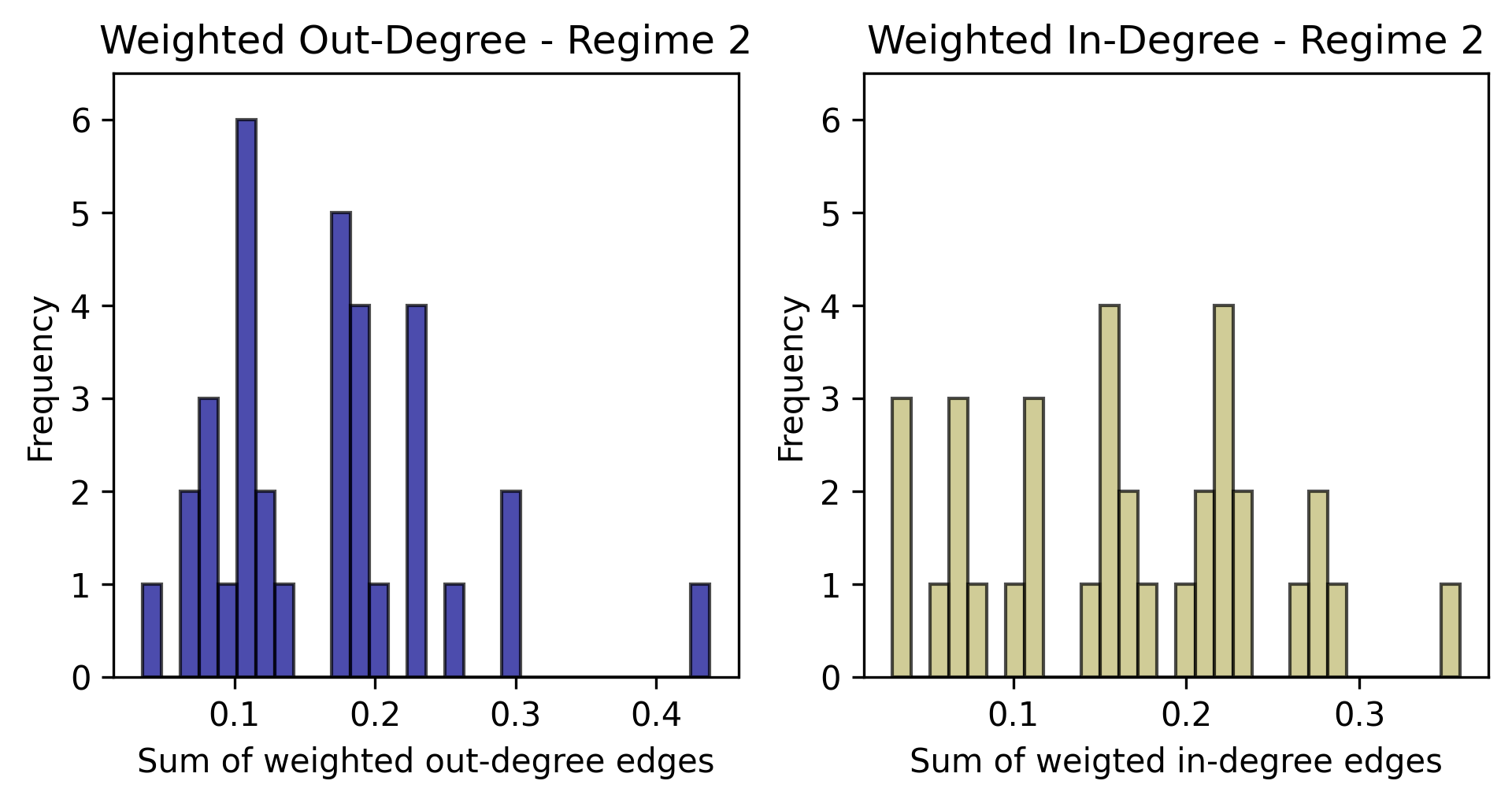}
            \caption{2021-08 to 2022-10}
        \end{subfigure}
        \hfill
        \begin{subfigure}[b]{0.32\textwidth}
            \includegraphics[width=\linewidth]{pics/degree_dis_regime3.png}
            \caption{After 2022-10}
        \end{subfigure}
        \caption{Weighted in-degree and out-degree distribution(social media).} \caption*{\small Notes: This figure shows the weighted in-degree and out-degree distributions for the news sentiment networks under different regimes.}
        \label{degreedistri_twi}
    \end{figure}

However, there is a slightly different story for social media information spillover. The network density fluctuates but within a narrow range. First of all, there is no evident shift during the second period as we visualise the information transmission spillover path in Figure \ref{regimes_twi}. We can further quantify this in the weighted in-degree and out-degree plot in Figure \ref{degreedistri_twi}. The weighted in/out-degree distribution is quite spread out across three regimes, with fewer companies with high in/out weighted degrees, suggesting their role in leading information spillover. We identify these companies in Table \ref{tab:twi_centrality_by_regime}. 

Some Magnificent Seven companies appear to have high centrality across all three regimes, such as AAPL, MSFT, NVDA, AMZN, GOOGL, and META, which suggests the noise traders or irrational traders, who make investment decisions based on their emotions, tend to have more discussions on those companies. These companies are often in the spotlight of the news and are perceived to have a significant impact on the whole industry. Notably, Advanced Micro Devices (AMD) is a leading global semiconductor company, and it has a high out-degree centrality through three regimes. In the meantime, we see more software and semiconductor companies play an important role in either transferring or receiving information in social media interactions. However, it is hard to distinguish the major changes between regime 1 and regime 2, while the information flow decreases a bit in regime 3. 


\begin{table}[H]
\caption{In-degree and Out-degree Centrality by Regime (Social Media)}
\label{tab:twi_centrality_by_regime}
\begin{tabularx}{\textwidth}{lXXl}
\toprule
\textbf{Regime 1} & & & \\
\midrule
\textbf{Company} & \textbf{In-degree} & \textbf{Company} & \textbf{Out-degree} \\
CRM (SW) & 0.359& AMD(SC)& 0.292
\\
V (FS)& 0.273 & INTC(SC) & 0.289 \\
MCO(CM)& 0.258&CRM(SW)& 0.267\\
\textcolor{red}{MSFT(SW)}&0.241&\textcolor{red}{GOOGL (CS)}&0.245\\
\textcolor{red}{AMZN (CC)}&0.215&MCHP(SC)&0.243\\
\midrule
\textbf{Regime 2} & & & \\
\midrule
\textbf{Company} & \textbf{In-degree} & \textbf{Company} & \textbf{Out-degree} \\
FIS(FS)& 0.358& AMD(SC)&0.438
\\
AMD(SC) & 0.286 & \textcolor{red}{NVDA(SC)} & 0.301\\
AMAT(SC)&0.280&MCO(CM)&0.293\\
\textcolor{red}{AMZN (CC)}&0.272&MCHP(SC)&0.259\\
ADP(PS)&0.261&\textcolor{red}{AAPL(TH)}&0.229\\
\midrule
\textbf{Regime 3} & & & \\
\midrule
\textbf{Company} & \textbf{In-degree} & \textbf{Company} & \textbf{Out-degree} \\
ORCL(SW) & 0.337& LRCX(SC)& 0.291
\\
CRM (SW)& 0.222 & \textcolor{red}{META (CS)}&0.253\\
ADBE(SW)&0.209&CRM(SW)&0.232\\
\textcolor{red}{META (CS)}&0.198&AMD(SC)&0.225\\
NOW(SW)&0.195&AMAT(SC)&0.215\\
\bottomrule
\end{tabularx}
\small{Notes: This table displays the top five nodes which have relatively high centrality under different regimes. The Magnificent Seven companies are highlighted in red to clearly indicate their positions in each regime. }
\end{table}

\begin{figure}[H]
        \centering
    \begin{subfigure}[b]{0.32\textwidth}
        \includegraphics[width=\linewidth]{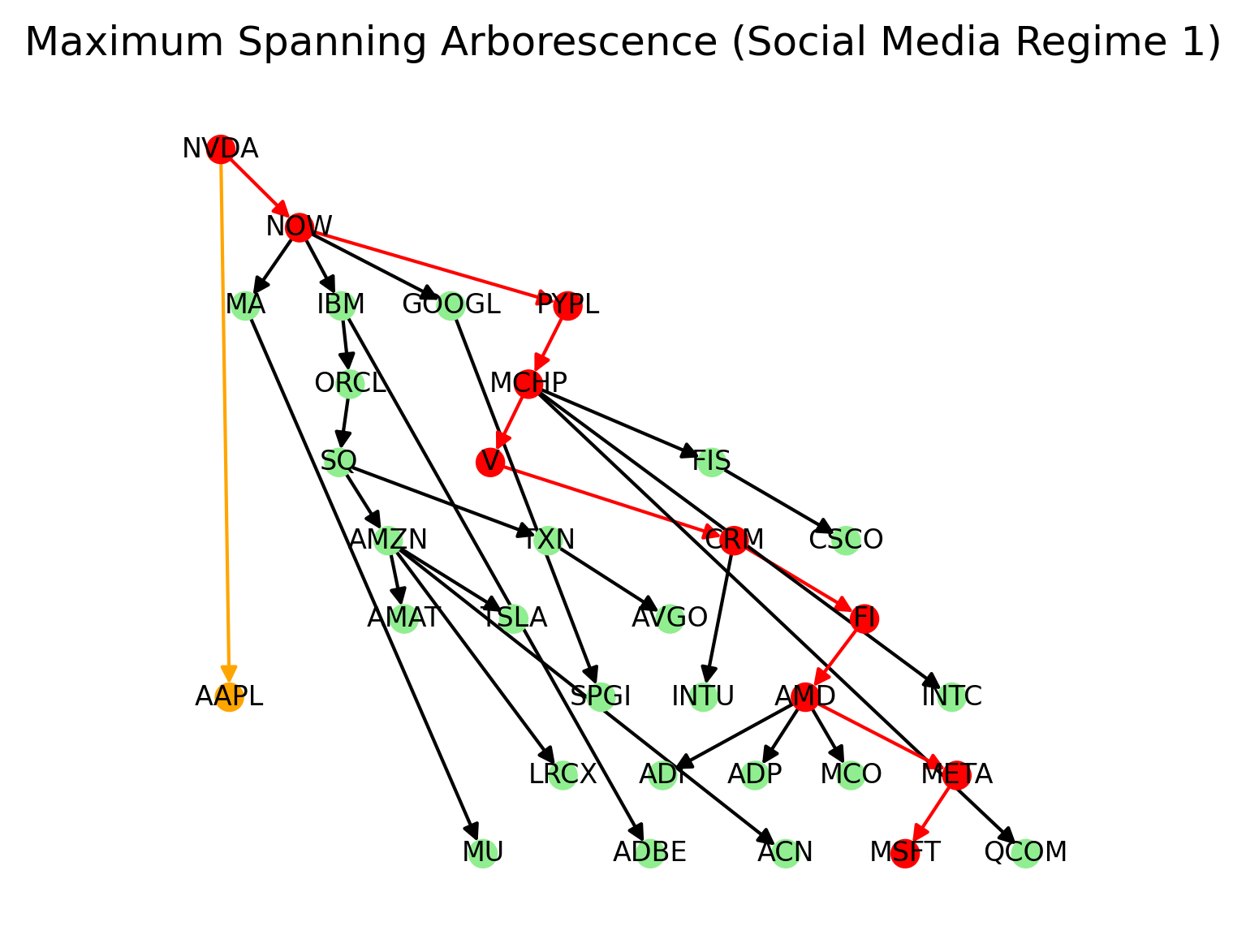}
            \caption{Before 2021-08}
        \end{subfigure}
        \hfill
        \begin{subfigure}[b]{0.32\textwidth}
            \includegraphics[width=\linewidth]{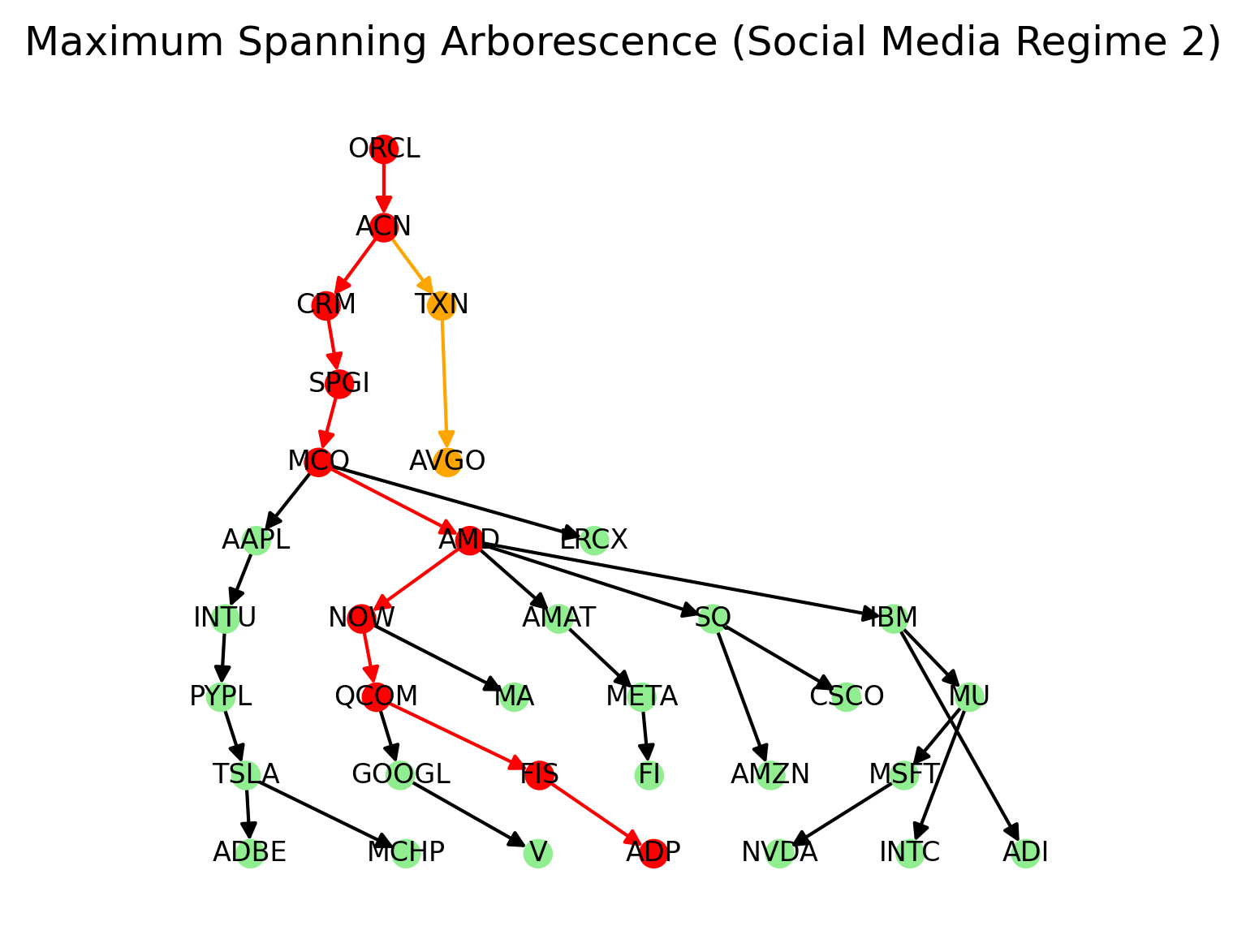}
            \caption{2021-08 to 2022-10}
        \end{subfigure}
        \hfill
        \begin{subfigure}[b]{0.32\textwidth}
            \includegraphics[width=\linewidth]{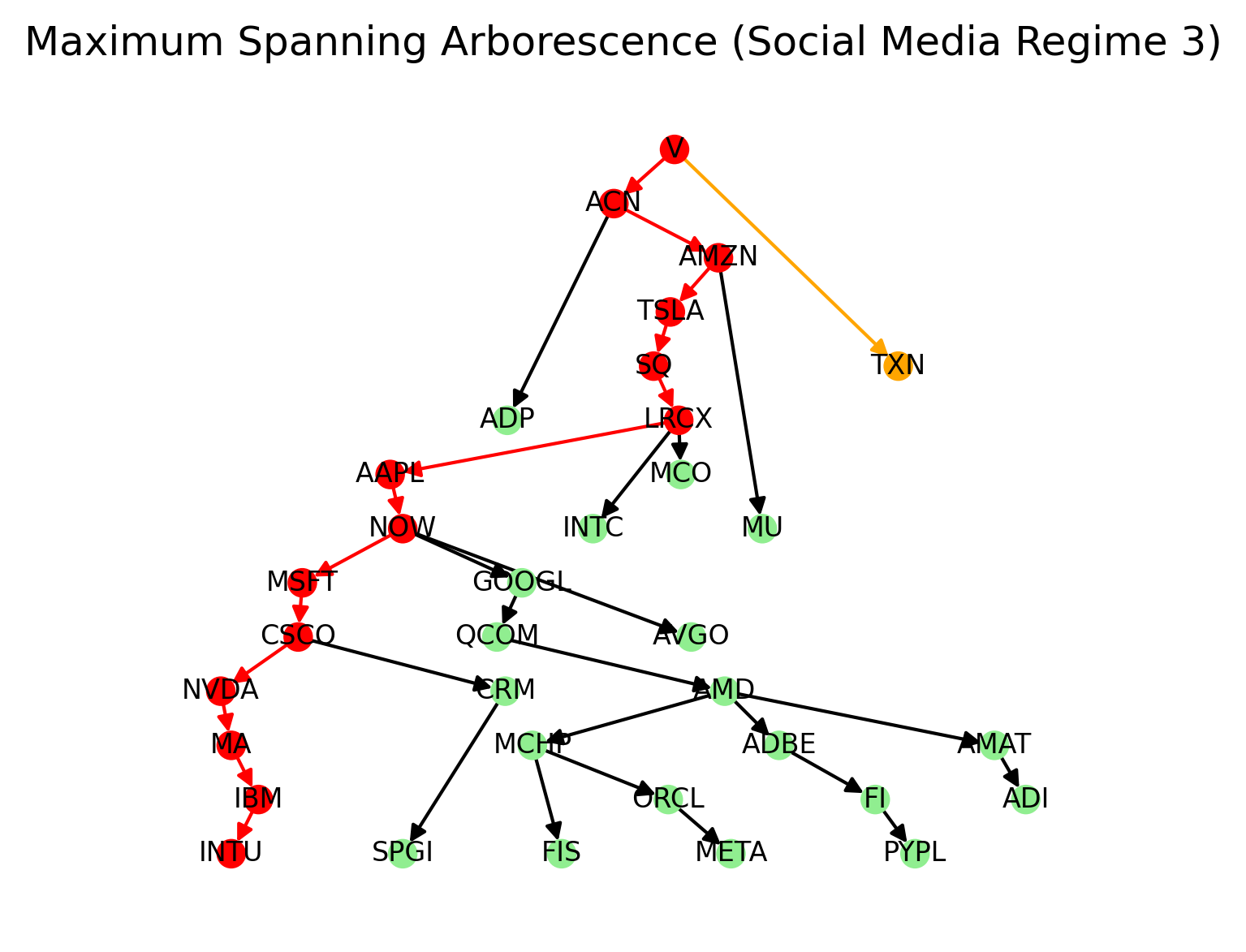}
            \caption{After 2022-10}
        \end{subfigure}
        \caption{Maximum Spanning Arborescence (Social Media).} \caption*{\small Notes: The maximum sum weights of the path are highlighted in red,  suggesting the strongest information flow. The minimum sum weights of the path are highlighted in orange. }
        \label{MSA social media}
    \end{figure}
\begin{figure}[H]
    \centering
    \begin{subfigure}{0.49\linewidth}
        \centering
        \includegraphics[width=\linewidth]{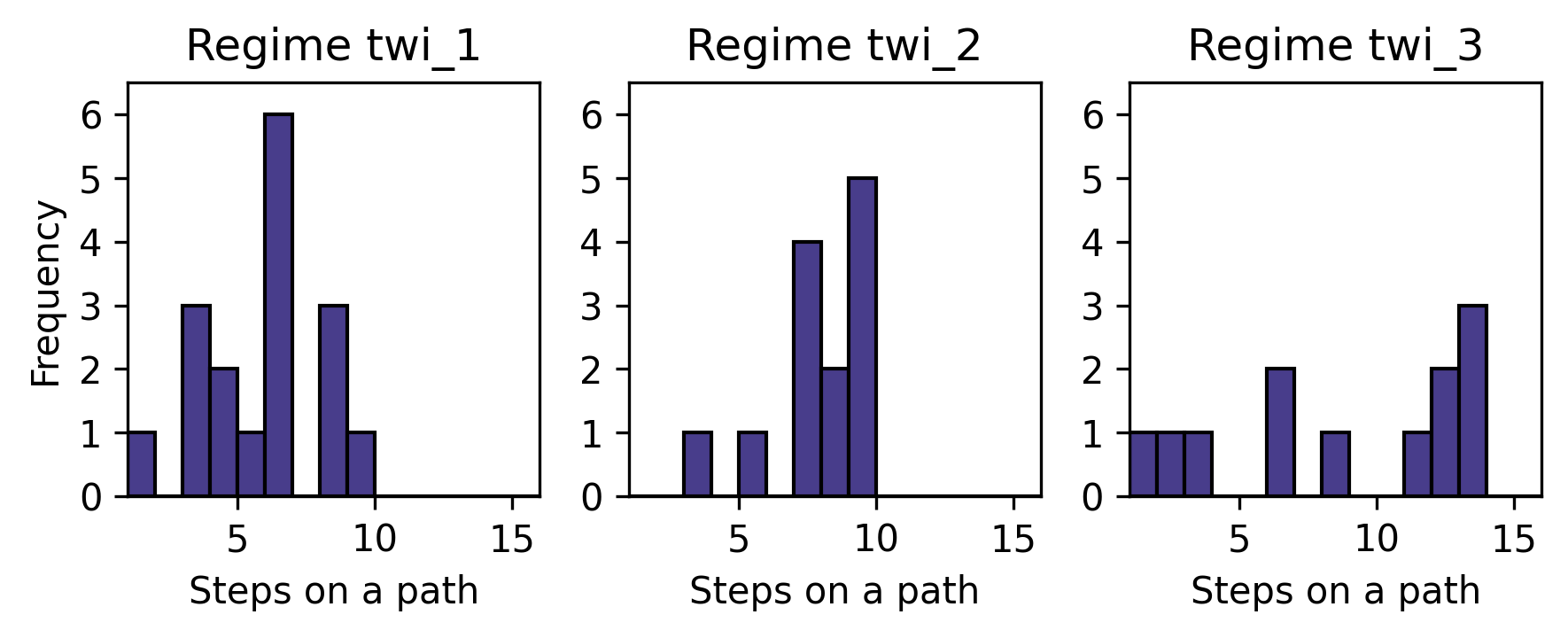}
        \caption{Number of steps}
    \label{fig:twi_steps_distri}
    \end{subfigure}
    \hfill
    \begin{subfigure}{0.49\linewidth}
        \centering
        \includegraphics[width=\linewidth]{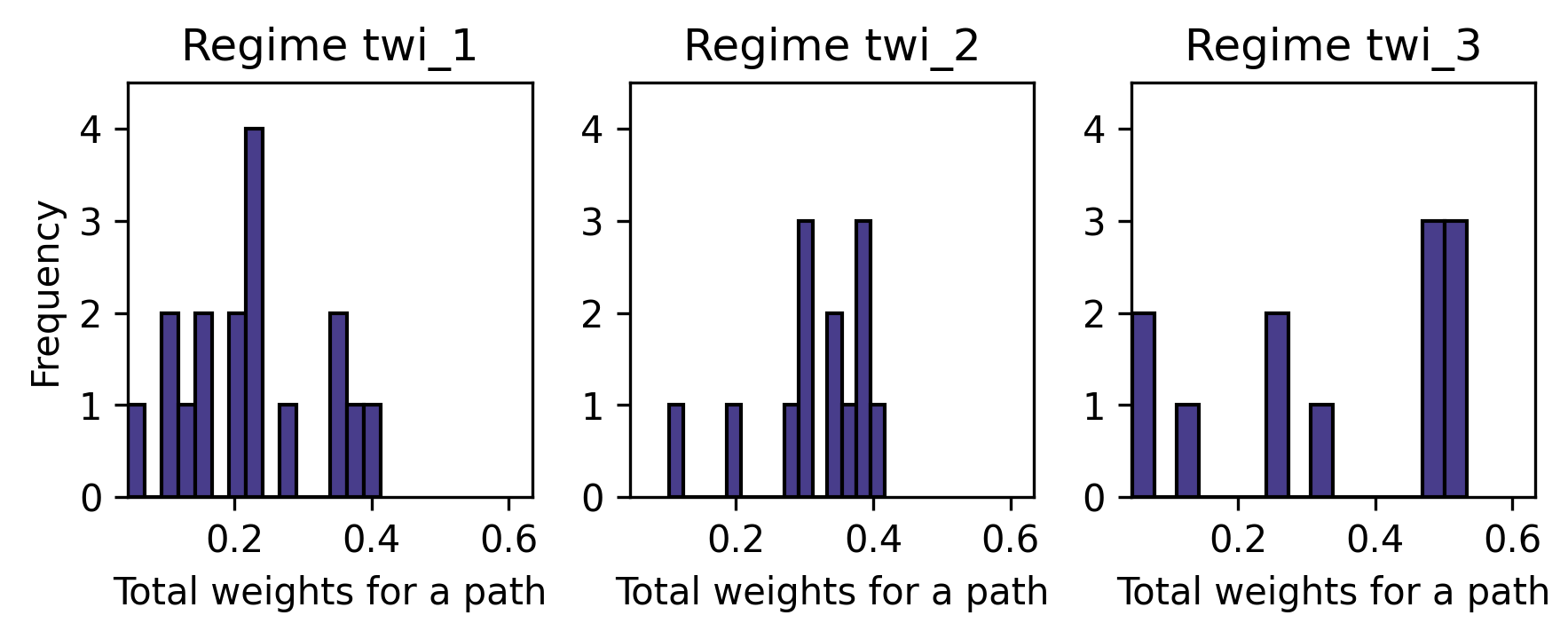}
        \caption{Total weights}
        \label{fig:twi_length_dis}
    \end{subfigure}
    \caption{Distribution of the number of steps and total weights on a path.}
\end{figure}

Same as news, we also obtain the maximum spanning arborescence for social media under three regimes. The visualisations are shown in Figure \ref{MSA social media}. The number of paths for each regime is 17, 13, and 12, respectively. The distribution of steps and the sum of weights on a path is plotted in Figure \ref{fig:twi_steps_distri} and Figure \ref{fig:twi_length_dis}. From the distribution plot, we can see that the average steps on a path increase from regime 1 to regime 3, indicating that the information takes longer to travel with slower information flow in the system, gradually. The average intensity of the information flow is 0.0408, 0.0421 and 0.0407, respectively, without evident difference, which again suggests the speculative and collective trait of social media. Companies such as NOW, NVDA, V, CRM, AMD, MSFT and ACN often appear on the path which has the maximum weights.  

In addition, we remark on several common findings from the news and social media spillover networks. Most of the Magnificent Seven companies not only play a significant role in the tech industry and financial market, but they also have high visibility in both news and social media. Some of the leading financial service companies, such as FIS, MA, as well as software and IT service companies like IBM, CRM, and NOW, provide transaction, payment and savings services, and also operate infrastructure for tech companies. It is not surprising that they are affected by external information and capable of spreading it.

\section{Conclusion}\label{conclude}
To conclude, this study has proposed sentiment spillover networks based on the transfer entropy method and has demonstrated their ability to capture the information flow between company pairs. It has given a comprehensive analysis of modelling sentiment spillover among the selected US technology companies through two different major channels of financial market information: news media and social media. Unlike previous studies that only focused on one information source, we compare both. We have analysed the dynamic sentiment spillover network pattern over time. The result signifies that the densities of the news and social media sentiment spillover networks do not always tell the same story. There is a clear spike in news-based information splliover network at the end of 2021, consistent with \cite{mbarki_sentiment_2022}.

However, we don't observe the same pattern in a social media-based spillover network. One of the reasons could be the disagreement on social media, as it collects all available information, which may cancel out any positive, negative and extreme attitudes. In contrast, news publishers selectively publish more authoritative information. In addition, our results indicate that the spillover in the social media information network has decreased since Twitter's rebranding to X. We also draw attention to companies with persistent roles in spreading and receiving the information over time in news media and social media by creating the heatmaps of the weighted in and out degrees over time, from which we can signify several important information hub companies, such as IBM, AVGO, TSLA and some financial services companies and semiconductor companies. 

Subsequently, we focus on three different regimes over the study period to have a deeper view of how the information transmission scheme among the companies has shifted. The information network visualisations, feathered with company degree distribution, clearly indicate the changes across the regimes and the difference in news and social media. Some of our findings align with the existing literature: for instance, IBM plays an important role in the news spillover network \citep{nyakurukwa_investor_2025}. But we spot some other interesting findings, for instance, we highlight that TSLA has a persistent role in spreading the information in social media over time. AVGO holds the most important position in news spillover networks. Additionally, by further extracting the maximum spanning arborescence, we notice different information spreading patterns under each regime, and find several magnificent seven companies, financial service companies, and companies that provide infrastructure to technology companies stand out as the major companies in bridging the information flow. These results provide guidelines to the investor to pay attention to the companies that are sensitive to the market news, and provide regulators with suggestions in cutting down the potential information contagion path by scrutinising market information in a timely manner among the companies when some crashes happen. 

We want to acknowledge some of the limitations in this study. First of all, the computational cost for the transfer entropy and the use of the bootstrap technique on a large selection of companies is quite intensive, even with parallel computing. We believe there is a better and faster way to compute the entries of the transfer entropy network. Second, we have not touched on comparing how positive, neutral and negative sentiment may spillover differently and how the disagreement on social media may affect the information spillover effect. Third, we also need to think about the use of the sentiment network, whether it can be an indicator for financial risk contagion, for example. 

Our study also opens several future research directions. One can adopt a similar modelling technique to explore and compare different information spillovers in different industries or countries. Second, researchers could look into whether the news spillover network is leading the social media spillover network or vice versa, and how this dynamic lead-lag relationship might change over time. In addition, the framework proposed can be potentially used as a measure of system aggregated sentiment index, which can serve as another layer of valuable information in asset pricing or market risk prediction. 

\newpage
\appendix
\section{Autoregressive Model}\label{app: AR}

\begin{figure}[H]
    \centering
    \includegraphics[width=\linewidth]{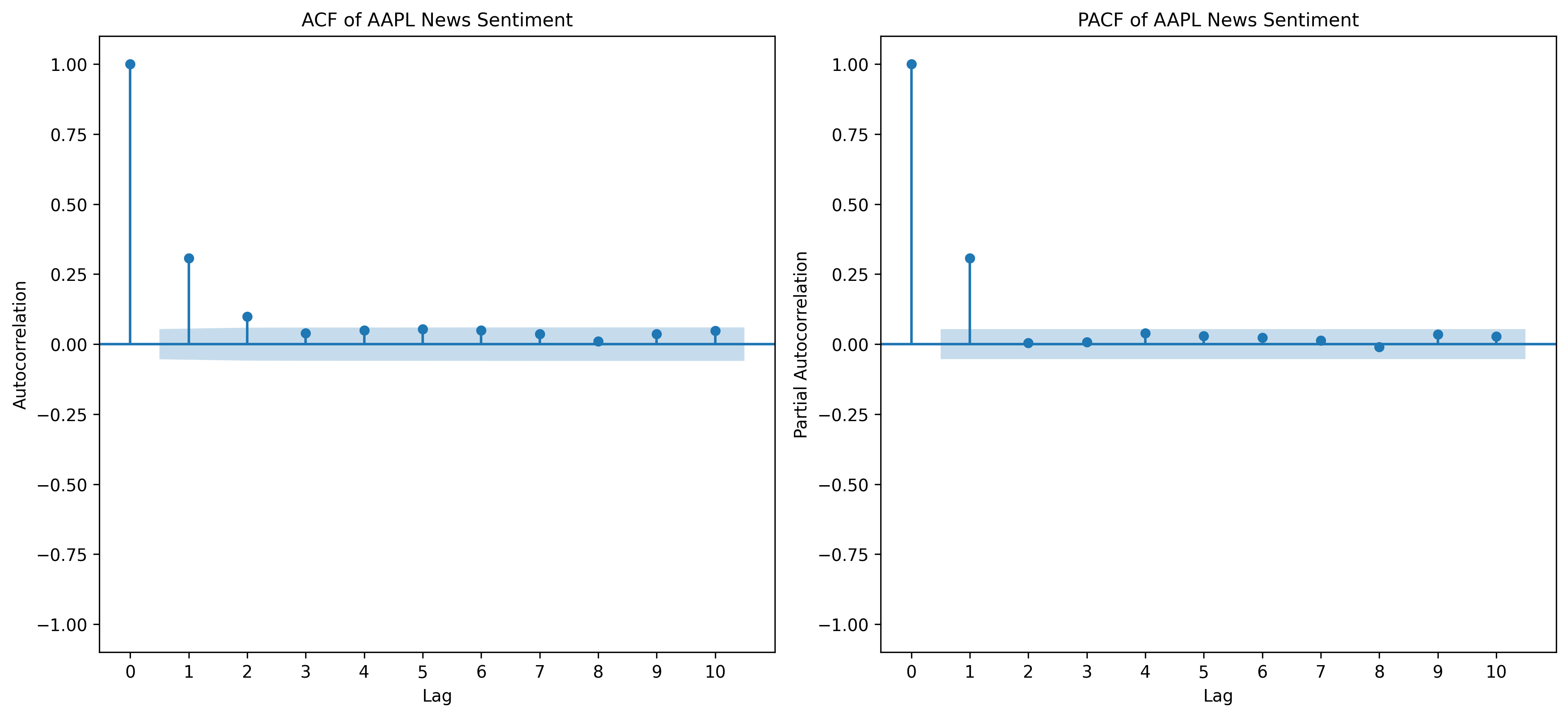}
    \caption{ACF and PACF plot of News sentiment series (AAPL as an example)}
    \label{fig:aaplpacf}
\end{figure}

\begin{figure}[H]
    \centering
    \includegraphics[width=\linewidth]{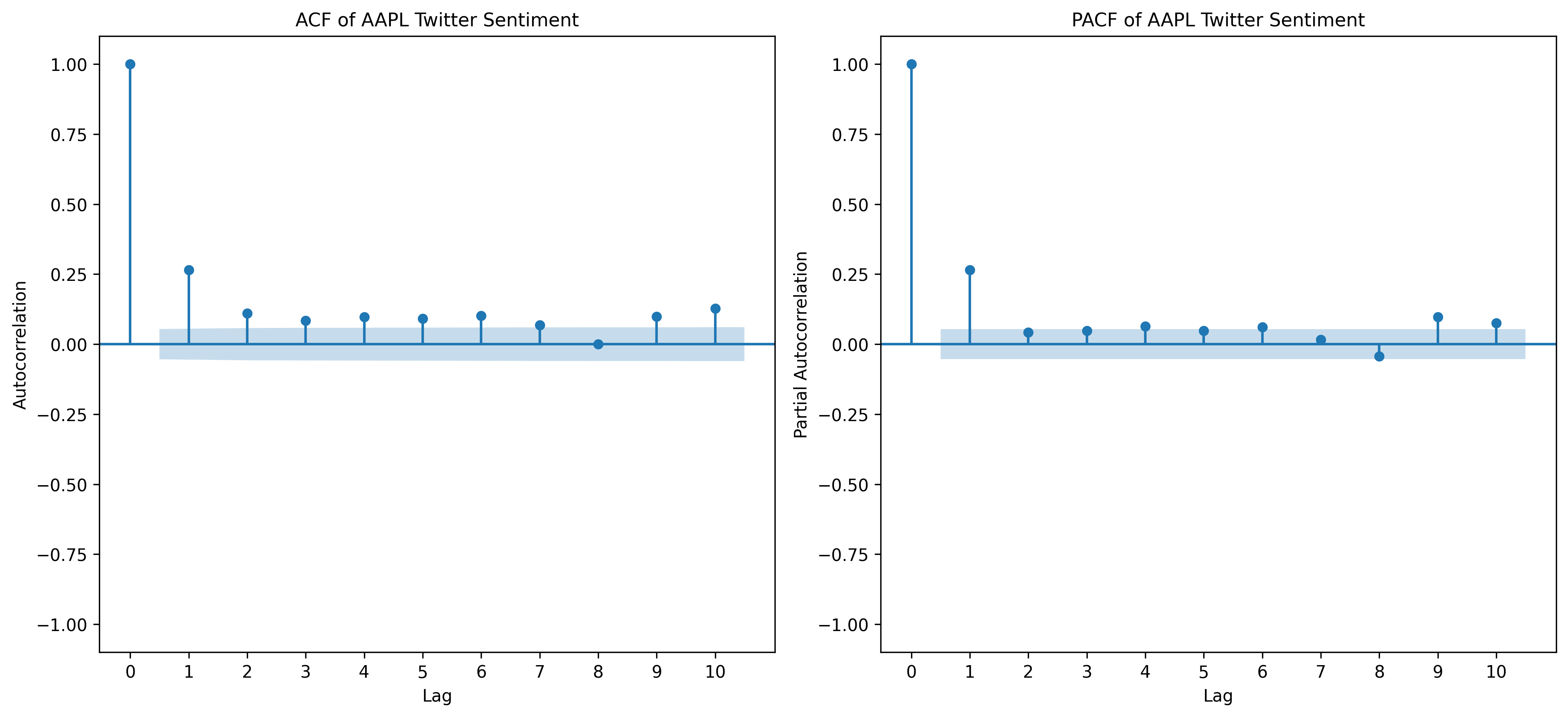}
    \caption{ACF and PACF plot of Social Media sentiment series (AAPL as an example)}
    \label{fig:aaplpacftwi}
\end{figure}

Preliminary time-series diagnostics based on the autocorrelation (ACF) and partial autocorrelation function (PACF) indicate that the sentiment series show statistically significant dependence at lag one in PACF plot and they tail off after lag $1$ in ACF plot (See the example of AAPL sentiment plot in Figure \ref{fig:aaplpacf} and \ref{fig:aaplpacftwi}). The structure is consistent with the autoregressive process of order one AR(1). Thus, for a given sentiment time series $x_t$, we consider the following autoregressive specification: 

\begin{equation}\label{app:ar1}
    x_t = o + \psi_1x_{t-1} + \epsilon_t, \hspace{1cm} \epsilon_t \sim \mathcal{N}(0,\sigma^2)
\end{equation}

We can use the available data to fit the AR(1) model and get the standard deviation ($\sigma$) for the error term ($\epsilon_t$) from the properties of AR(1) process: 
\begin{equation}
    \text{Var}(x_t) = \frac{\sigma^2}{1-\psi_1^2},
\end{equation}
which is derived by taking the variances on both sides of the equation \eqref{app:ar1}:

\begin{equation}
\begin{aligned}
\text{Var}(x_t)=&\text{Var}(o)+\psi_1^2\text{Var}(x_{t-1})+\text{Var}(\epsilon_t)\\
=&\psi_1^2 \text{Var}(x_{t-1})+\sigma^2
   \end{aligned} 
\end{equation}
where $\text{Var}(x_t)=\text{Var}(x_{t-1})$ based on stationary assumption. The estimated standard deviation ($\sigma$) for the error term is then used in the exponential decay model to impute the missing sentiment values. 
\section{The list of technology companies}\label{appen: company lists}
The following table (Table \ref{appen:companylists}) presents key information on the technology companies selected for analysis, including their stock tickers and GICS classification.
\begin{table}[H]
\caption{List of technology companies, tickers, and GICS code.}
\tiny
\setlength{\tabcolsep}{2mm}
\resizebox{\linewidth}{!}{\begin{tabular}{lll}
            \toprule
            \textbf{Companies}    & \textbf{Tickers}     & \textbf{GICS Classification (Abbrv.)} \\
            \midrule
Apple Inc.    & AAPL & Technology Hardware (TH)\\
Microsoft Corp.        & MSFT    & Software (SW) \\
NVIDIA Corp. & NVDA        & Semiconductors (SC)\\
Alphabet Inc  &GOOGL&Communication Services\\
Tesla Inc&TSLA&Consumer Cyclical\\
Meta Platforms Inc&META&Communication Services\\
Amazon.com Inc& AMZN &Consumer Cyclical\\
\hline
\hline
Visa Inc. & V    & Financial Services (FS) \\
Mastercard & MA & Financial Services (FS)\\
Adobe Inc. & ADBE & Software (SW)\\
PayPal Holding Inc. & PYPL & Financial Services (FS)\\
Salesforce Inc. & CRM        & Software (SW)\\
Cisco System Inc. & CSCO & Communications Equipment (CE)      \\
Intel Corp.  & INTC    & Semiconductors (SC)\\
Broadcom Inc. & AVGO        & Semiconductors (SC)\\
Oracle Corp. & ORCL        & Software (SW)\\
Accenture PLC & ACN    & IT Services (IT)\\
Advanced Micro Devices Inc. & AMD        & Semiconductors (SC)\\
Texas Instruments Inc. & TXN        & Semiconductors (SC)\\
QUALCOMM Inc. & QCOM        & Semiconductors (SC)\\
Intuit Inc. & INTU        & Software (SW)\\
ServiceNow Inc. & NOW        & Software (SW)\\
International Business Machines Corp. & IBM        & IT Services (IT)\\
S\&P Global Inc. & SPGI        & Capital Markets (CM)\\
Applied Materials Inc. & AMAT & Semiconductors (SC)\\
Automatic Data Processing Inc. & ADP & Professional Services (PS)\\
Analog Devices Inc. & ADI & Semiconductors (SC)\\
Lam Research Corp. & LRCX & Semiconductors (SC)\\
Block Inc. & SQ (XYZ now) & Financial Services (FS)\\
Micron Technology Inc. & MU & Semiconductors (SC)\\
Moody's Corp. & MCO & Capital Markets (CM)\\
Fiserv Inc. & FI & Financial Services (FS)\\
Fidelity National Information Services Inc. & FIS & Financial Services (FS)\\
Microchip Technology Inc. & MCHP & Semiconductors (SC)\\
            \bottomrule
        \end{tabular}}

\footnotesize{Notes: This table provides information on the selected 34 tech companies.}
\label{appen:companylists}
\end{table}

\bibliographystyle{apalike}

\end{document}